\documentclass[final,5p,times,twocolumn]{elsarticle}

\usepackage{amsmath,amssymb,amsfonts}
\usepackage{algorithmic}
\usepackage{algorithm}
\usepackage{array}
\usepackage{textcomp}
\usepackage{stfloats}
\usepackage{url}
\usepackage{verbatim}
\usepackage{graphicx}
\usepackage{multirow}
\usepackage{subfigure}
\usepackage{threeparttable}
\usepackage{float}
\usepackage{booktabs}
\usepackage{notoccite}
\usepackage{longtable}
\usepackage{makecell}
\usepackage{lineno}
\journal{Arxiv}

\begin{document}

\begin{frontmatter}

\title{Exploring EEG and Eye Movement Fusion for Multi-Class Target RSVP-BCI}

\author[label1,label2]{Xujin Li \fnmark[1]}
\author[label1]{Wei Wei \fnmark[1]}
\author[label1,label3]{Kun Zhao}
\author[label1,label3]{Jiayu Mao}
\author[label1,label2]{Yizhuo Lu}
\author[label1,label3]{Shuang Qiu \corref{cor1}}
\author[label1,label2,label3]{Huiguang He \corref{cor1}} 
\cortext[cor1]{Corresponding authors: Huiguang He, Shuang Qiu}
\ead{huiguang.he@ia.ac.cn}
\fntext[fn0]{The first two authors contributed equally to this work.}

\affiliation[label1]{organization={Key Laboratory of Brain Cognition and Brain-inspired Intelligence Technology, Institute of Automation, Chinese Academy of Sciences},%Department and Organization 
            city={Beijing},
            postcode={100190},  
            country={China}}

\affiliation[label2]{organization={School of Future Technology, University of Chinese Academy of Sciences (UCAS)},%Department and Organization 
            city={Beijing},
            postcode={100049}, 
            country={China}}

\affiliation[label3]{organization={School of Artificial Intelligence, University of Chinese Academy of Sciences (UCAS)},%Department and Organization 
            city={Beijing},
            postcode={100049}, 
            country={China}}

\begin{abstract}
    Rapid Serial Visual Presentation (RSVP)-based Brain-Computer Interfaces (BCIs) facilitate high-throughput target image detection by identifying event-related potentials (ERPs) evoked in electroencephalography (EEG) signals. The RSVP-BCI systems effectively detect single-class targets within a stream of images but have limited applicability in real-world scenarios that require detecting multiple target categories. Multi-class RSVP-BCI systems address this limitation by simultaneously identifying the presence of a target and distinguishing its category. However, existing multi-class RSVP decoding algorithms predominantly rely on single-modality EEG decoding, which restricts their performance improvement due to the high similarity between ERPs evoked by different target categories. In this work, we introduce eye movement (EM) modality into multi-class RSVP decoding and explore EEG and EM fusion to enhance decoding performance. First, we design three independent multi-class target RSVP tasks and build an open-source dataset comprising EEG and EM signals from 43 subjects. Then, we propose the \textbf{M}ulti-class \textbf{T}arget \textbf{R}SVP \textbf{E}EG and \textbf{E}M fusion \textbf{Net}work (MTREE-Net) to enhance multi-class RSVP decoding. Specifically, a dual-complementary module is proposed to strengthen the differentiation of uni-modal features across categories. To improve multi-modal fusion performance, we adopt a dynamic reweighting fusion strategy guided by theoretically derived modality contribution ratios. Furthermore, we reduce the misclassification of non-target samples through knowledge transfer between two hierarchical classifiers. Extensive experiments demonstrate the feasibility of integrating EM signals into multi-class RSVP decoding and highlight the superior performance of MTREE-Net compared to existing RSVP decoding methods. The proposed MTREE-Net and open-source dataset provide a promising framework for developing practical multi-class RSVP-BCI systems.
    
\end{abstract}


\end{frontmatter}


\section{Introduction}
    Rapid Serial Visual Presentation (RSVP)-based Brain-Computer Interfaces (BCIs) utilize electroencephalography (EEG) signals to efficiently and rapidly detect rare target events within a stream of images \cite{lees2018review}.  The RSVP-BCI systems have gained significant attention for their potential to enhance human-computer interaction and collaboration \cite{marathe2015effect,lin2018novel}, and have been applied in various fields including target image detection \cite{cecotti2010convolutional,alpert2013spatiotemporal,marathe2015improved}, spellers \cite{acqualagna2013gaze}, face recognition \cite{chen2024eeg}, and anomaly detection \cite{barngrover2015brain}.

    In the RSVP paradigm, visual stimuli are presented sequentially at high speed in the same spatial location \cite{lees2018review}. Subjects are required to identify images containing specific target items from other non-target images, where target images are rare and distributed sparsely and randomly within the stimulus sequence. These rare target images evoke Event-Related Potentials (ERPs) that include the P300 component, which is a prominent positive peak occurring around 200-600 ms after target stimulus onset \cite{squires1976effect,polich2007updating}. Over the past decade, numerous EEG decoding methods have been developed to detect these evoked ERPs for target image detection, which significantly enhance RSVP decoding performance \cite{cecotti2010convolutional,EEGNet,PLNet,PPNN,ji2024novel}. Additionally, many studies have shown that participants exhibit eye movement (EM) responses, such as changes in pupil area and fixation locations, in response to target images in RSVP tasks \cite{privitera2010pupil,luo2023erp}. This is because EM signals are closely associated with cognitive processes in visual tasks such as visual search \cite{najemnik2005optimal} and target detection \cite{qian2009decision}. Recent studies have introduced EM data into RSVP-EEG decoding, which demonstrates the effectiveness of integrating the EM modality in RSVP tasks \cite{luo2023erp,mao2023cross}. Thus, both EEG and EM signals provide discriminative information that can improve RSVP decoding, and their integration has the potential to further enhance fusion performance.
    
\begin{figure}[!htbp]
        \centering
        \subfigure[Single-class target RSVP-BCI (Target: people)]{
			\includegraphics[width=0.95\linewidth]{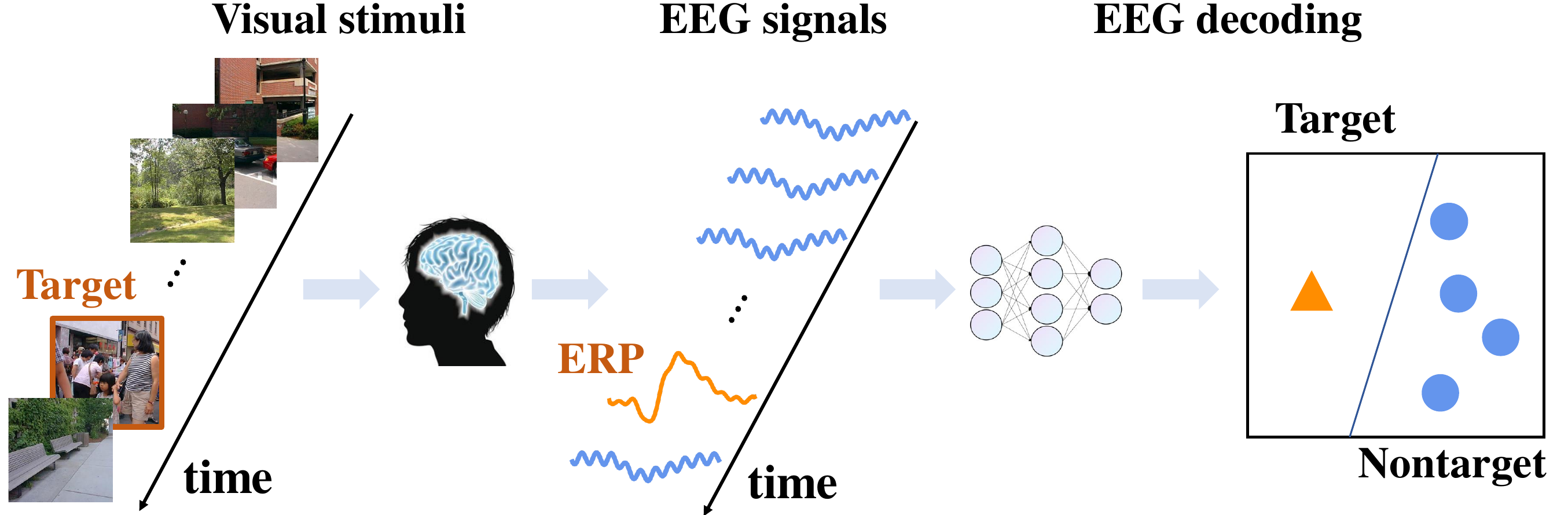}
		}
		\subfigure[Multi-class target RSVP-BCI (Target-1: people, Target-2: car)]{
			\includegraphics[width=0.95\linewidth]{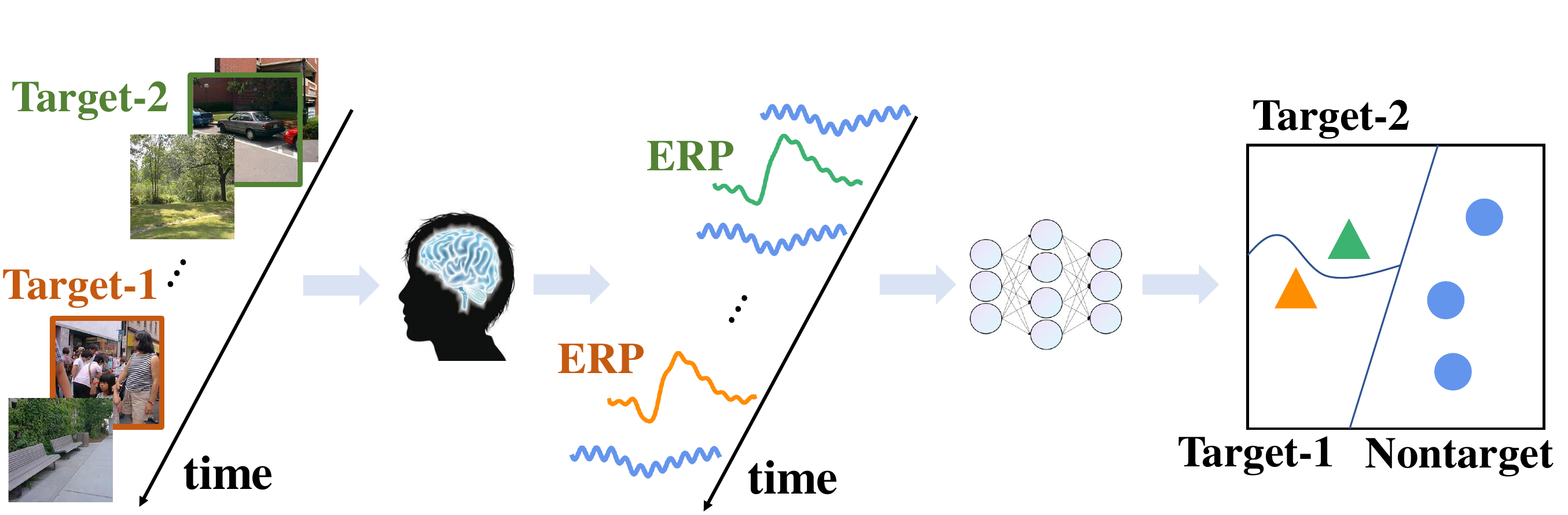}
		}
		\caption{The diagrams of (a) single-class target RSVP-BCI and (b) multi-class target RSVP-BCI. In the single-class target RSVP-BCI, subjects identify target images (e.g., people) within an image sequence. The decoding model detects target presence by identifying ERPs evoked by the target images. In the multi-class target RSVP-BCI, participants identify multiple target categories (e.g., people and cars) within the image sequence. The decoding model detects the presence of targets and differentiates between target categories by identifying and classifying ERPs evoked by different target classes.}
        \label{multi-class RSVP}
    \end{figure}

    The advancements in RSVP decoding methods have substantially improved decoding performance, establishing the RSVP-BCI system as a robust and efficient tool for target image detection \cite{zhang2023uav}. However, traditional RSVP-BCI systems are restricted to single-class target detection, which is a binary classification task that distinguishes between ERPs induced by target images and harmonic signals induced by non-target images \cite{PPNN} (see Fig. \ref{multi-class RSVP}(a)). This limits BCI systems to recognizing only one target category within an image sequence. With the expansion of application scenarios, the demand for multi-class target detection in practical applications has increased \cite{marathe2015effect,cecotti2012multiclass,zhou2024multiband}. Consequently, research on multi-class target RSVP paradigms is essential for developing RSVP-BCI systems that can simultaneously detect target existence and identify their specific categories. 

    In recent years, researchers have extended the traditional single-class target RSVP paradigm to the multi-class target RSVP paradigm. As shown in Fig. \ref{multi-class RSVP}, compared to single-class target detection, multi-class target detection distinguishes targets from non-targets and differentiates between multiple target categories. In 2022, Wang et al. designed an RSVP task requiring participants to identify images containing people (target-1) or cars (target-2) from street scenarios \cite{wang2023review} (see Fig. \ref{multi-class RSVP}(b)). In 2024, Wei et al. designed four multi-class target RSVP tasks and demonstrated that distinct target categories within a single task evoke P300 responses with no significant statistical differences in latency but with significant differences in amplitude \cite{wei2024preliminary}. Thus, multi-class target detection can be realized by distinguishing between target-specific P300 signals and non-target responses. The multi-class RSVP decoding extends beyond single-class target detection by distinguishing ERPs across different target categories while maintaining discrimination between target and non-target signals (see Fig. \ref{multi-class RSVP}). Based on recent advances in the multi-class target RSVP paradigm, Chen et al. (2024) introduced the common representation extraction targeted stacked convolutional auto-encoder for detecting two targets alongside a non-target in RSVP tasks \cite{chen2024cre}. In the same year, Wei et al. also proposed the multi-scale dilated convolutions network (MDCNet) for multi-class RSVP decoding \cite{wei2024preliminary}. However, current research on multi-class RSVP decoding remains in its preliminary stages, and the performance of existing methods is insufficient to meet the requirements of practical applications. On the one hand, the high similarity among target-evoked ERPs in EEG signals poses a significant challenge for accurate multi-class RSVP decoding. On the other hand, existing multi-class RSVP decoding methods are limited to uni-modal EEG decoding and ignore the complementary potential of EM signals that also exhibit distinct responses to target stimuli in RSVP tasks \cite{luo2023erp,mao2023cross}. This reliance on single-modality decoding may restrict further improvements in decoding performance.

    Therefore, we aim to enhance the multi-class RSVP decoding performance by exploring EEG and EM fusion to improve the distinguishability of different categories of target and non-target evoked responses. Effective EEG and EM fusion in multi-class RSVP decoding is a challenging task in terms of datasets and algorithms. For the dataset, no research has yet incorporated the EM modality into multi-class RSVP decoding, resulting in a lack of high-quality datasets to support algorithm development. For the algorithm, first, the varying classification capabilities of EEG and EM signals can hinder uni-modal training and reduce the differentiation of uni-modal features across different classes \cite{peng2022balanced}. Second, many existing algorithms employ reweighting methods for effective EEG and EM feature fusion \cite{mao2023cross,fu2024cross}. However, the optimization of the reweighting modules in these methods relies solely on the final classification loss without explicit guidance from the discriminative capabilities of each modality, which may hinder the effective weighting of modalities based on their importance to classification. Finally, current fusion algorithms fail to leverage the unique characteristics of multi-class RSVP data, limiting their ability to enhance the model’s discriminability across different classes in multi-class RSVP decoding.
    
    In this study, we first design three independent multi-class target RSVP tasks and build an open-source dataset from 43 subjects containing both EEG and EM signals. Then we propose the Multi-class Target RSVP EEG and EM fusion Network (MTREE-Net) to enhance multi-class RSVP decoding. In MTREE-Net, We employ multi-scale convolutions to extract both local and broad temporal features from EEG signals \cite{wang2021performance}, and utilize a single-layer convolution to efficiently capture temporal patterns in EM signals \cite{mao2023cross}. Then, we propose a dual-complementary module that enhances cross-modal feature complementarity to improve the discriminative capabilities of uni-modal features. Subsequently, by theoretically analyzing each modality's contribution to the classification results and deriving corresponding contribution ratios to guide the optimization of the reweighting module, we employ a dynamic weighting strategy that adaptively adjusts the importance of each modality during feature fusion. Finally, focusing on the characteristics of multi-class RSVP data, we propose a hierarchical self-distillation module comprising a binary classifier for target versus non-target discrimination and a triplet classifier for multi-class target classification. The knowledge between these two hierarchical classifier is transferred to reduce non-target misclassification rates. The main contributions of this work are summarized as follows:

    \begin{itemize}
        \item[1)] We design and conduct three independent multi-class target RSVP tasks and collect EEG and EM data from 43 subjects simultaneously, which is the first multi-class target RSVP dataset containing both EEG and EM signals. Our self-collection dataset (DOI:10.57760/sciencedb.17705) is open source at https://doi.org/10.57760/sciencedb.17705.

        \item[2)] We propose the Multi-class Target RSVP EEG and EM fusion Network (MTREE-Net) to enhance multi-class RSVP decoding. To the best of our knowledge, this is the first model that fuses EEG and EM signals for multi-class RSVP decoding.

        \item[3)] We propose a contribution-guided reweighting module to enhance multi-modal fusion performance by introducing the contribution ratio of each modality to guide the reweighting module optimization. Additionally, a hierarchical self-distillation module is proposed to further reduce the misclassification of non-target samples.

        \item[4)] Extensive experiments are conducted on our open-source dataset. The experiments demonstrate the feasibility of utilizing the EM modality in multi-class RSVP tasks and the excellent performance of our proposed model in multi-class RSVP decoding. The source codes have been submitted as supplementary material and will be released upon acceptance.
    \end{itemize}

\section{Related Work}
    This section provides a brief review of previous studies related to our study, including single-class target RSVP-EEG decoding methods, multi-class target RSVP-EEG decoding methods, and EEG and EM signals fusion methods.
    
\subsection{Single-Class Target RSVP-EEG Decoding Methods}
    Over the past decade, RSVP decoding research has primarily focused on single-class target RSVP tasks. Meanwhile, deep learning approaches \cite{lecun2015deep} have become increasingly prevalent and achieved notable performance in EEG signal classification. Cecotti et al. first introduced CNN for P300 detection in 2011, which included multiple model variants and proved the effectiveness of CNN for RSVP-EEG decoding \cite{cecotti2010convolutional}. In 2018, Lawhern et al. (2018) introduced EEGNet, a compact CNN using depthwise separable convolution \cite{chollet2017xception} to reduce model parameters while maintaining performance \cite{EEGNet}. In 2021, Zang et al. proposed the Phase-Locked Network (PLNet), which leveraged the phase-locked characteristic of ERPs by applying spatial convolution across different time periods to extract spatiotemporal features \cite{PLNet}. In the same year, Wang et al. proposed MS-CNN, a CNN with multi-scale structures containing three kernels of different sizes, and integrating the feature maps by the three temporal filters to capture more discriminative features \cite{wang2021performance}. In 2022, Li et al. proposed a Phase Preservation Neural Network (PPNN) to learn phase information of EEG signals in all frequency bands and improve classification performance in the RSVP task \cite{PPNN}. These CNN-based methods have demonstrated superior RSVP-EEG decoding performance. Over the past two years, researchers have investigated Transformer architectures for EEG decoding. Song et al. (2023) proposed the EEG-conformer, a compact convolutional Transformer designed to integrate local and global features within a single EEG classification framework \cite{song2022eeg}. In 2023, Luo et al. introduced a multi-view fusion model (STSTNet) based on Transformers for EEG-based visual recognition, which combined features from both the spatiotemporal and spectral-temporal domains of EEG signals \cite{luo2023dual}. In 2024, Ji et al. developed a hybrid decoding model called HCANN, which employed depthwise separable convolutions to separate the temporal dependencies of EEG signals and utilized a multi-head self-attention mechanism to capture spatial activation patterns for RSVP decoding \cite{ji2024novel}. After training on sufficient EEG data, these Transformer-based methods demonstrate excellent performance in RSVP decoding \cite{li2025temporal}.

\subsection{Multi-Class Target RSVP-EEG Decoding Methods}
    In the past two years, researchers have extended the RSVP task to include multi-class target scenarios and have investigated decoding algorithms for these tasks \cite{wu2022review, wang2023review}. In 2024, Chen et al. proposed a novel classification model based on a stacked convolutional auto-encoder for two-target RSVP-EEG classification \cite{chen2024cre}. This model generated a common representation for each target class, which reduced variability across different trials and improved the distinction between the EEG signals evoked by two-class targets. In the same year, Wei et al. introduced a Multi-scale Dilated Convolution Network (MDCNet) for the same task \cite{wei2024preliminary}. MDCNet employed multi-scale dilated convolutions to extract comprehensive information about EEG amplitude and phase in the temporal dimension, as well as channel relationships in the spatial dimension. These studies on multi-class target RSVP decoding are still in the early stages, and further improvements in model performance are necessary to satisfy the practical application.
    
\subsection{EEG and Eye Movement Signals Fusion Methods}
    With the increasing application of multi-modal fusion in BCI tasks, EM signals are recognized as playing a significant role in enhancing performance. This leads to growing attention on integrating EEG and EM data to improve BCI system efficacy \cite{abiri2019comprehensive,luo2023erp,wang2024research}. Emotion recognition is an area where EEG and EM fusion have been extensively applied. For example, Zhao et al. (2021) introduced a modified Dense Co-Attention Symmetric Network (mDCAN) to combine information from EEG and EM modalities. This approach retains modality-specific information at each level by integrating outputs from each DCAN layer to increase the effectiveness of multi-modal fusion \cite{nguyen2018improved,zhao2021multimodal}. In 2024, Fu et al. proposed a dual-branch fusion network (FGFRNet) to fuse EEG and EM signals. This network includes a feature guidance module that uses EEG features to direct the extraction of eye movement features, along with a feature reweighting module to identify emotion-related features within EM signals \cite{fu2024cross}. In RSVP decoding, Mao et al. (2023) developed a Cross-Modal Guiding and Fusion Network (CMGFNet) to fully utilize EEG and EM modalities and fuse them for better RSVP decoding performance \cite{mao2023cross}. On the one hand, these algorithms use reweighting methods for EEG and EM feature fusion, but their optimization relies only on final classification loss, ignoring the discriminative capabilities of each modality. On the other hand, these methods achieved excellent performance in their corresponding tasks, but their effectiveness in multi-class RSVP decoding is limited due to a lack of adaptation to the specific data characteristics in multi-class RSVP decoding.

\section{Materials}
    We design and implement three independent multi-class target RSVP tasks to collect EEG and EM data, and establish the ``NeuBCI Multi-Class Target Detection RSVP EEG and EM Dataset." For convenience, the three multi-class target RSVP tasks are denoted as Task A, Task B, and Task C, respectively. The target categories for each task are non-civil and civil aircraft in Task A, storage tanks and centers in Task B, and harbors and parking lots in Task C. The ``NeuBCI Multi-Class Target Detection RSVP EEG and EM Dataset'' is available at https://doi.org/10.57760/sciencedb.17705.
    
\subsection{Subjects}
    We recruit 43 subjects (mean age: $23.8\pm2.4$ years) to participate in the multi-class target RSVP experiment, 24 of whom are female. All participants have normal or corrected-to-normal vision, and 42 are right-handed. None of the participants report any history of visual disorders, neurological diseases, or injuries. Before participation, each participant provides written informed consent. The experimental procedures are approved by the Institutional Review Board of the Institute of Automation, Chinese Academy of Sciences.

\subsection{Multi-Class Target RSVP Paradigm}
    \begin{figure}[ht]
        \centering
        \subfigure[]{
			\includegraphics[width=0.95\linewidth]{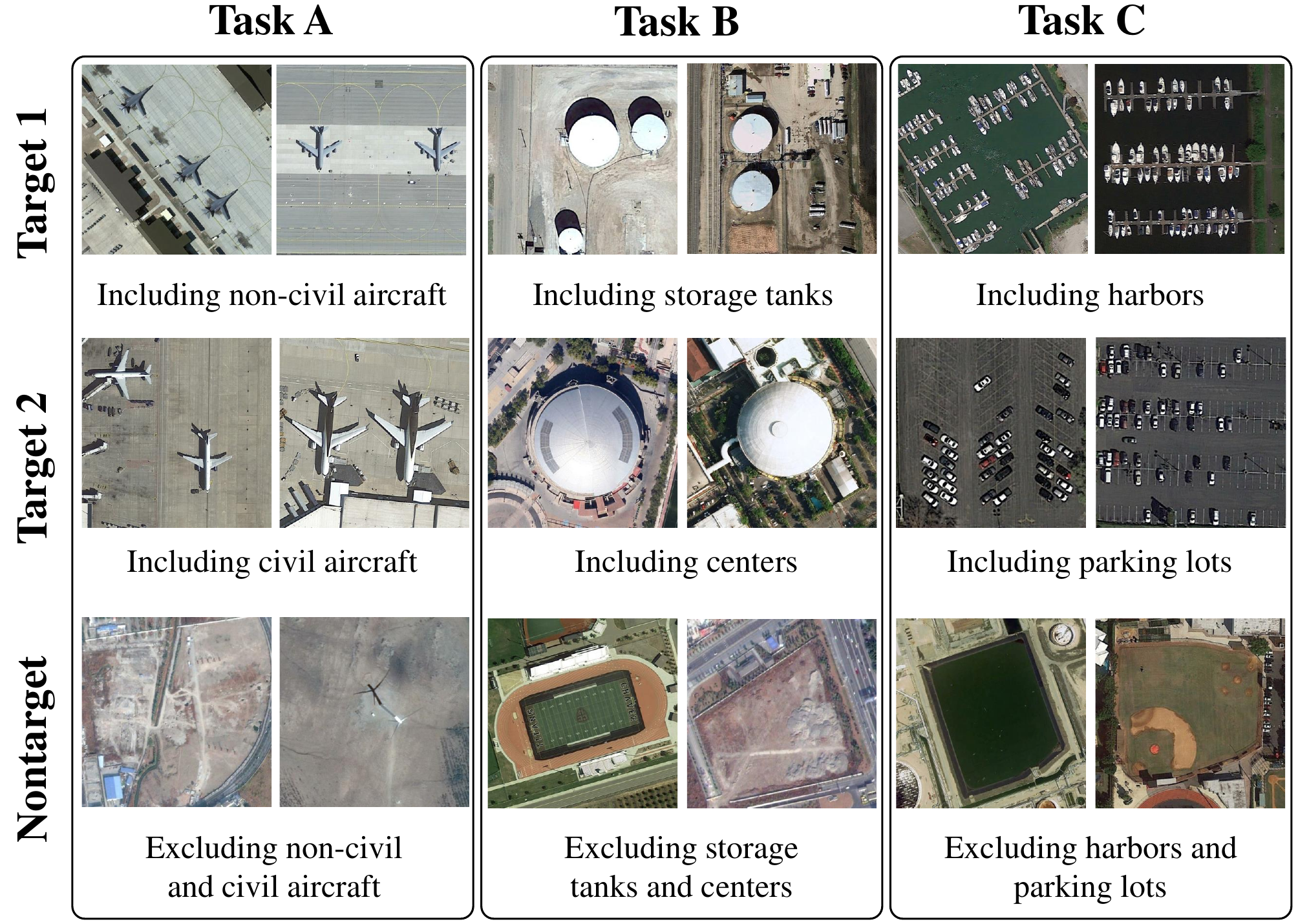}
		}
		\subfigure[]{
			\includegraphics[width=0.93\linewidth]{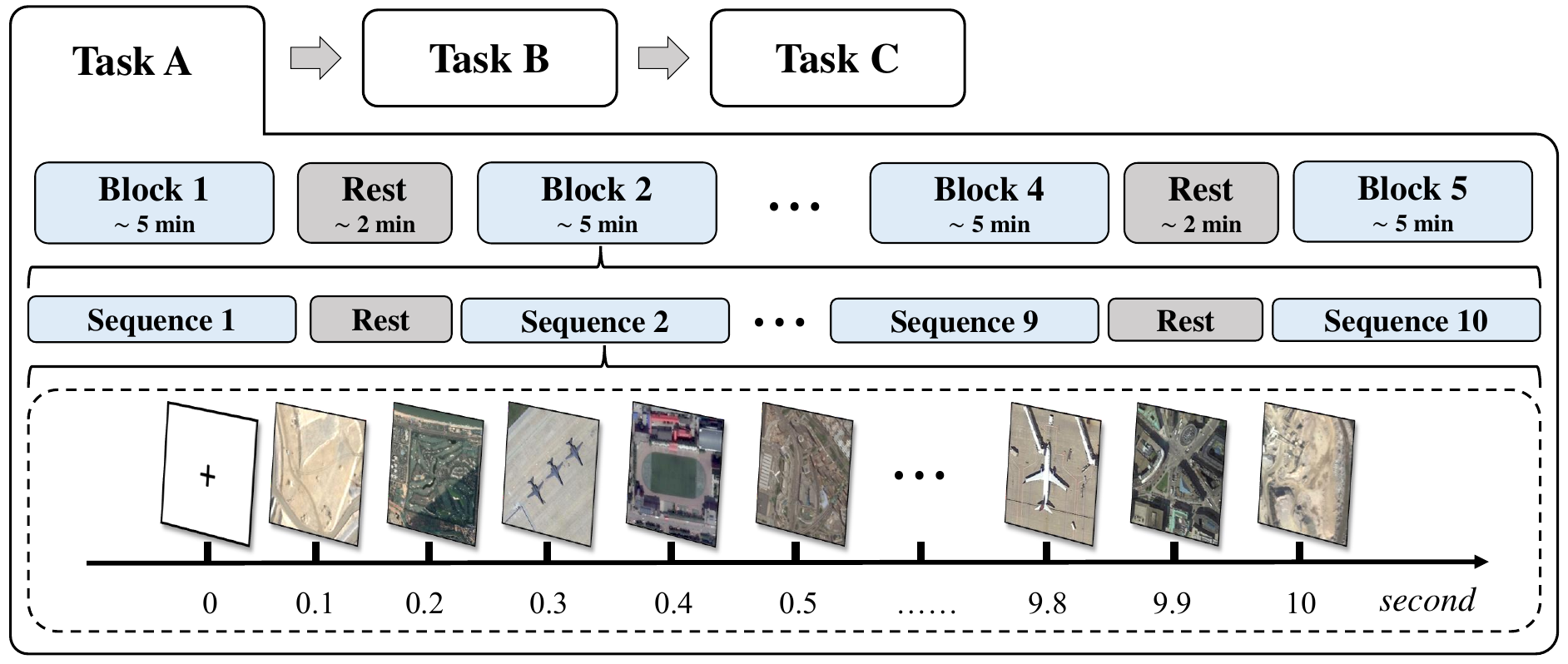}
		}
		\caption{Illustration of the RSVP paradigm. (a) Examples of target-1, target-2, and non-target images in Task A, Task B, and Task C. The stimulus images in three tasks are sourced from the remote sensing Dior dataset \cite{li2020object}. (b) Experimental settings about the division of tasks, blocks, and sequences for each subject.}
        \label{RSVP experiment}
    \end{figure}
    
    In the experiment, we design three independent multi-class target RSVP experiments for practical applications. We utilize the remote sensing Dior dataset \cite{li2020object} to simulate the application of remote sensing observation. Two-class objects with similar visual characteristics are chosen as target-1 and target-2 for each task: non-civil and civil aircraft (Task A), storage tanks and centers (Task B), and harbors and parking lots (Task C). In Task A, both non-civil and civil aircraft are types of aircraft. In Task B, storage tanks and centers are circular or elliptical objects. In Task C, harbors and parking lots are areas containing numerous small objects that are regularly arranged. 

    As a result, we have designed three independent RSVP tasks: (1) Task A requires participants to identify both non-civil and civil aircraft from remote sensing images; (2) Task B focuses on identifying both storage tanks and centers in remote sensing images; (3) Task C involves retrieving both harbors and parking lots in remote sensing images. Within each task, the stimulus images containing the two task-specified target classes are designated as target-1 and target-2 images, while the remaining images are considered non-target images. The two target classes in each task do not co-occur within the same image. Figure \ref{RSVP experiment} (a) illustrates a few representative examples of the stimulus images.

    As shown in Fig. \ref{RSVP experiment} (b), all subjects complete Task A, B, and C sequentially, where each task consists of one practice block followed by five formal blocks. The practice block familiarizes participants with the task structure, while formal blocks are used for data collection. Each block comprises 10 sequences, with each sequence containing 100 images. The intervals between sequences are controlled by the participants. Within each sequence, images are presented randomly at 10 Hz. After achieving a relaxed state, participants are presented with a sequence of images and instructed to identify the target-1 and target-2 images. The occurrence probability of target images (target-1 and target-2) is 6 $\%$, and the number of target-1 and target-2 images are around the same. Visual stimuli are displayed on a 17-inch monitor (1280 × 1024 pixels) with individual images sized at 500 × 500 pixels. Participants are seated 1 meter from the monitor. The luminance and contrast are set at 50 during stimulus presentation. All screen settings remain fixed throughout the experiment.

\subsection{Data Acquisition and Preprocessing}
    The data acquisition and preprocessing are consistent across all three RSVP tasks. The specific processes are as follows:
    
\subsubsection{Data Acquisition}
    The EEG data are recorded using a SynAmp2 Amplifier (NeuroScan, Australia) with a 64-channel Ag/AgCl electrode cap following the international 10-20 system. The electrode impedances are maintained below 10 $k\Omega$, with AFz serving as the ground electrode and the vertex as the reference. Data are sampled at 1000 Hz. Both EEG and eye movement signals are recorded simultaneously during the experiment. Bilateral eye movements including pupil pixel area, horizontal (X), and vertical (Y) gaze coordinates are recorded using an EyeLink 1000 plus system (SR Research) at the same sampling rate of 1000 Hz. A chin rest is used to minimize head movement while maintaining attention on the target images. Before data collection, eye tracking is calibrated using a 9-point grid, with maximum and average errors below $1^{\circ}$ and $0.5^{\circ}$, respectively, which ensures stable eye movement recordings.

    \begin{figure*}[htbp]
		\centering
		\includegraphics[width=\linewidth]{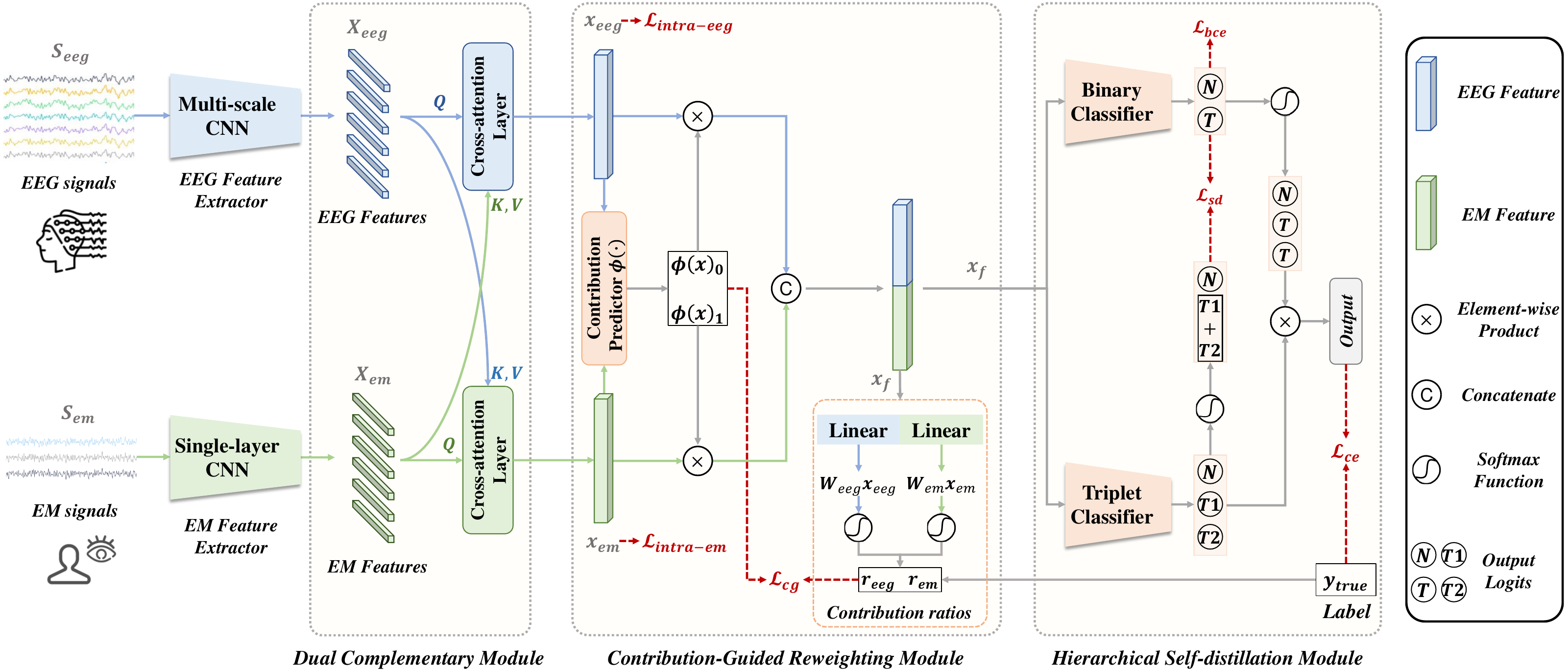}
		\caption{The structure of our proposed MTREE-Net model consists of a two-stream feature extractor, a dual-complementary module (DCM), a contribution-guided reweighting module (CG-RM), and a hierarchical self-distillation module (HSM). The network employs several loss functions: intra-modal triplet cross-entropy losses ($\mathcal{L}_{intra\mbox{-}eeg}$ and $\mathcal{L}_{intra\mbox{-}em}$), binary cross-entropy loss ($\mathcal{L}_{bce}$), triplet cross-entropy loss ($\mathcal{L}_{ce}$), contribution-guided loss with $L_1$-norm ($\mathcal{L}_{cg}$), and self-distillation loss using symmetric Kullback-Leibler divergence ($\mathcal{L}_{sd}$). The output logits $N$, $T$, $T1$, and $T2$ correspond to non-target, target, target-1, and target-2 classes, respectively.}
        \label{model framework}
    \end{figure*}

\subsubsection{Data Preprocessing}
    The preprocessing stage has two main parts. Initially, both EEG and eye movement (EM) signals are down-sampled to 128 Hz, with EEG data additionally filtered through a third-order Butterworth bandpass filter (0.5-15 Hz). Subsequently, the continuous EEG and EM data are segmented into trials from the onset of the presented image to 1000 ms after the onset, and the EEG data of -200-0 ms are used for baseline correction. The processed dataset contains 5000 paired EEG-EM samples per subject in each task, where the number of target-1 samples and target-2 samples are both around 300, and the remaining samples are non-target. The results in this article rely on single-trial EEG and EM modality sample pairs.

\section{Methods}
    This section introduces the Multi-class Target RSVP EEG and EM fusion Network (MTREE-Net) designed for multi-class target RSVP decoding. As illustrated in Fig. \ref{model framework}, MTREE-Net consists of four main components: a two-stream feature extractor, a dual-complementary module (DCM), a contribution-guided reweighting module (CG-RM), and a hierarchical self-distillation module (HSM). The inputs to the MTREE-Net are raw EEG signals and EM signals. First, a two-stream feature extractor extracts multi-scale features from EEG signals while compressing features from various EM signal components. The DCM then enhances the distinction between uni-modal features of two modalities across different classes by leveraging cross-attention mechanisms to achieve modal complementarity. Subsequently, the features of both modalities are weighted and fused in the CG-RM, which is guided by the contribution ratios of each modality to classification results for optimization. Finally, the HSM specially designed for multi-class target RSVP data further reduces misclassification of non-target samples through self-distillation between binary (target and non-target) and triplet (target-1, target-2, and non-target) classification outputs.

\subsection{Feature Extractor}
    The feature extractor in MTREE-Net is a two-stream architecture including an EEG feature extractor and an EM feature extractor which extracts features from EEG and EM signals respectively. The inputs to the feature extractor are raw EEG signals ($\boldsymbol{S}_{eeg}\in \mathbb{R}^{C_{eeg}\times T}$) and EM signals ($\boldsymbol{S}_{em}\in \mathbb{R}^{C_{em}\times T}$), where $C_{eeg}$ represents the number of EEG channels, $C_{em}$ denotes the number of EM components, and $T$ denotes the number of sampling points. 

    EEG signals comprise transient and oscillatory patterns of varying temporal durations which reflect continuous brain activity, making multi-scale analysis particularly effective \cite{santamaria2020eeg}. This approach enables simultaneous capture of both fine-grained local patterns and broader temporal dependencies \cite{wang2021performance}. Hence, we employ a multi-scale convolutional neural network for EEG feature extraction. In contrast, EM signals exhibit simpler temporal patterns primarily consisting of basic features like pupil size and fixation position \cite{mao2023cross}. We implement a single-layer convolutional network that can sufficiently extract discriminative features while maintaining computational efficiency. Table \ref{structure of feature extractor} presents the detailed architecture of the feature extractor.

    \setlength{\tabcolsep}{2.0mm}{
    \begin{table}[htbp]
        \footnotesize
        \centering
        \renewcommand\arraystretch{1.3}
        \caption{The architecture of the feature extractor.}
        \label{summary}
        \begin{tabular}{c|c|ccl}
            \toprule[1.2pt]
            \textbf{\makecell{Modality}} & \textbf{Block} & \multicolumn{1}{c}{\textbf{Layer}} & \textbf{Kernels} & \multicolumn{1}{c}{\textbf{Output}} \\
            \hline
            \multirow{16}{*}{\textbf{EEG}} & \multirow{5}{*}{\textbf{1}}& Input & -  & $(1, C_{eeg}, T)$ \\
            &  & \multirow{2}{*}{MS-Conv} &  \multirow{2}{*}{$\{(8, 1, \lfloor\frac{T}{2^{t}}\rfloor)\}_{t=1}^{4}$} & \multirow{2}{*}{$\{(8, C_{eeg}, T)\}_{t=1}^{4}$} \\
            &  &  &   &  \\
            & & BN+ELU &  - & $\{(8, C_{eeg}, T)\}_{t=1}^{4}$ \\
            & & Dropout &  $p=0.2$ & $\{(8, C_{eeg}, T)\}_{t=1}^{4}$ \\
            \cline{2-5}
            & \multirow{5}{*}{\textbf{2}}& DS-Conv &  $\{(16, C_{eeg}, 1)\}_{t=1}^{4}$  & $\{(16, 1, T)\}_{t=1}^{4}$ \\
            & & BN+ELU & -  & $\{(16, 1, T)\}_{t=1}^{4}$ \\
            & & Dropout   & $p=0.2$ & $\{(16, 1, T)\}_{t=1}^{4}$ \\
            & & Concat & - & $(64, 1, T)$ \\
            & & AvgPool & $(1, 4)$  & $(64, 1, \lfloor T/4\rfloor)$ \\
            \cline{2-5}
            & \multirow{6}{*}{\textbf{3}}& \multirow{2}{*}{MS-Conv} & \multirow{2}{*}{$\{(8, 1, \lfloor\frac{T}{2^{t}\times4}\rfloor)\}_{t=1}^{4}$}  & \multirow{2}{*}{$\{(8, 1, \lfloor T/4\rfloor)\}_{t=1}^{4}$} \\
            & &  &   &  \\
            & & BN+ELU  & -  & $\{(8, 1, \lfloor T/4\rfloor)\}_{t=1}^{4}$ \\
            & & Dropout   & $p=0.2$ & $\{(8, 1, \lfloor T/4\rfloor)\}_{t=1}^{4}$ \\
            & & Concat  & -  & $(32, 1, \lfloor T/4\rfloor)$ \\
            & & AvgPool & $(1, 2)$  & $(32, 1, \lfloor T/8\rfloor)$ \\
            \hline
            \multirow{4}{*}{\textbf{EM}} & \multirow{4}{*}{\textbf{1}}& Input & -  & $(1, C_{em}, T)$ \\
            &  & Conv & $(32, C_{em}, 8)$  & $(32, 1, \lfloor T/8\rfloor)$ \\
            & & Leaky Relu & - & $(32, 1, \lfloor T/8\rfloor)$ \\
            & & Dropout &  $p=0.2$ & $(32, 1, \lfloor T/8\rfloor)$ \\
            \bottomrule[1.2pt]
        \end{tabular}
        \begin{tablenotes}
            \item \footnotesize \emph{\textbf{Note:}} Both MS-Conv employ the same padding, while the Conv uses a stride of $(1,8)$. $\lfloor\cdot\rfloor$ denotes the floor function, which maps a real number to the largest integer less than or equal to it.
            \item \footnotesize \emph{\textbf{Abbreviations:}} MS-Conv: Multi-Scale Convolution layer; DS-Conv: Depthwise Separable Convolution layer; BN: Batch Normalization.
        \end{tablenotes}
        \label{structure of feature extractor}
    \end{table}

\subsubsection{Multi-Scale Convolution EEG Feature Extractor}
    The EEG feature extractor consists of three main blocks designed to capture both temporal and spatial characteristics of EEG signals effectively. Inspired by previous multi-scale convolution studies in RSVP task \cite{wei2024preliminary,santamaria2020eeg}, we implement multi-scale convolutions with kernel sizes ranging from $(1, T/2)$ to $(1, T/16)$, decreasing by a factor of 2. Initially, Block 1 applies multi-scale convolutions with kernel sizes $\{(8, 1, \lfloor\frac{T}{2^{t}}\rfloor)\}_{t=1}^{4}$ for multi-scale temporal feature extraction. Subsequently, Block 2 employs depthwise separable convolutions $\{(16, C_{eeg}, 1)\}_{t=1}^{4}$ \cite{chollet2017xception} to capture inter-channel relationships and compress spatial information while reducing computational complexity. The four groups of multi-scale features output by depthwise separable convolutions are concatenated and followed by the $(1,4)$ average pooling to reduce temporal dimension. Block 3 implements another set of multi-scale convolutions with kernels $\{(8, 1, \lfloor\frac{T}{2^{t}\times4}\rfloor)\}_{t=1}^{4}$ for broader temporal feature extraction. The output multi-scale features are concatenated and followed by the $(1,2)$ average pooling. Each convolution in the EEG feature extractor is followed by batch normalization and elu activation \cite{sharma2017activation}. Finally, the output of the EEG feature extractor is $\boldsymbol{X}_{eeg}\in \mathbb{R}^{c\times d}$, where $c=32$ and $d=16$. 

\subsubsection{Single-Layer Convolution EM Feature Extractor}
    The EM feature extractor only contains one convolution layer which consists of $c$ convolution kernels with a kernel size of $(C_{em},\lfloor T/d\rfloor)$ and the stride is set to the same value. The convolution kernel is set up in this way so that the EM feature and EEG feature sizes are the same to facilitate subsequent interaction and fusion. The activation function used in the model is the leaky relu \cite{xu2020reluplex} activation function. The output of the EM feature extractor is $\boldsymbol{X}_{em}\in \mathbb{R}^{c\times d}$, which is consistent with the dimensions of $\boldsymbol{X}_{eeg}$.

\subsection{Dual-Complementary Module}
    EEG signals provide richer multidimensional features (e.g., temporal and spatial domains) compared to EM signals, enabling higher discriminability between different stimuli in RSVP tasks \cite{mao2023cross}. This disparity introduces training imbalances in multi-modal training, which can potentially prevent the optimization of the EM stream \cite{peng2022balanced}. This optimization imbalance degrades both uni-modal EM decoding performance and multi-modal fusion effectiveness.

    To enhance the distinguishability of uni-modal features across different classes, we propose a dual-complementary module (DCM) based on the cross-attention mechanism \cite{xu2023multimodal}. The inputs of the DCM are EEG ($\boldsymbol{X}_{eeg}$) and EM ($\boldsymbol{X}_{em}$) features extracted by the feature extractor. The linear projection layer maps input features from both modalities into separate feature spaces and then both modalities mutually extract common features based on the similarity matrix. We utilize a two-head attention mechanism in our model. To concisely represent the inference process, we present the single-head attention mechanism as a representative example. The mathematical formulation is as follows:
    \begin{equation}
            \boldsymbol{X}_{eeg}^{DCM} = \boldsymbol{X}_{eeg} + \sigma(\frac{\boldsymbol{X}_{eeg}\boldsymbol{W}_{Q}^{(1)}{\boldsymbol{W}_{K}^{(1)}}^{T}\boldsymbol{X}_{em}^{T}}{\sqrt{d_{k}}})\boldsymbol{X}_{em}\boldsymbol{W}_{V}^{(1)},
            \label{EM to EEG}
    \end{equation}
    \begin{equation}
            \boldsymbol{X}_{em}^{DCM} = \boldsymbol{X}_{em} + \sigma(\frac{\boldsymbol{X}_{em}\boldsymbol{W}_{Q}^{(2)}{\boldsymbol{W}_{K}^{(2)}}^{T}\boldsymbol{X}_{eeg}^{T}}{\sqrt{d_{k}}})\boldsymbol{X}_{eeg}\boldsymbol{W}_{V}^{(2)},
            \label{EEG to EM}
    \end{equation}
    where $\boldsymbol{X}_{eeg}^{DCM} \in \mathbb{R}^{c\times d}$ and $\boldsymbol{X}_{em}^{DCM} \in \mathbb{R}^{c\times d}$ are the EEG and EM features output from the DCM respectively. The linear matrices $\boldsymbol{W}_{Q}^{(i)}\in \mathbb{R}^{d\times d}$, $\boldsymbol{W}_{K}^{(i)}\in \mathbb{R}^{d\times d}$, and $\boldsymbol{W}_{V}^{(i)}\in \mathbb{R}^{d\times d}$ ($i=1, 2$) are learnable parameter matrices. The $\sigma(\cdot)$ represents the softmax function.

\subsection{Contribution-Guided Reweighting Module}
    Following the DCM, we concatenate features ($\boldsymbol{X}_{eeg}^{DCM}$, $\boldsymbol{X}_{em}^{DCM}$) from both modalities for multi-modal fusion. While this straightforward concatenation applies equal weights to both modalities, such uniform weighting is suboptimal because of the varying discriminative characteristics of the two modalities. Previous research has attempted to address this limitation by implementing learnable reweighting modules that dynamically adjust the importance of each modality \cite{fu2024cross,mao2023cross}. However, these modules are optimized solely through classification loss without explicit information guidance. Consequently, this optimization approach without direct guidance can fail to properly weight modalities according to their discriminative capabilities.
    
    In the contribution-guided reweighting module (CG-RM), we first analyze each modality's contribution to classification performance in concatenation-based multi-modal fusion. Based on this analysis, we introduce explicit guidance based on modal contributions to guide the reweighting module's training, which can enhance fusion performance by increasing fusion weights of the modality with greater contribution to correct classification.

\subsubsection{Analysis of Modality Contribution}
    \medskip
    In CG-RM, the output of the DCM {\small $(\boldsymbol{X}_{eeg}^{DCM}$, $\boldsymbol{X}_{em}^{DCM})$} are first flattened as one-dimensional vectors ($\boldsymbol{x}_{eeg}, \boldsymbol{x}_{em} \in \mathbb{R}^{(c\times d)}$) for concatenation. After the features of the two modes are concatenated ($\boldsymbol{x} \in \mathbb{R}^{(2\times c\times d)}$), the logits output is obtained through the linear layer for final classification. For simplicity, we use $[\cdot,\cdot]$ to denote concatenation. Because the dual-class target RSVP decoding task is triplet classification,  $\boldsymbol{W} \in \mathbb{R}^{3\times(2\times c\times d)}$ and $b \in \mathbb{R}^{3}$ are used to denote the parameters of the linear classifier to produce the logits output:
    \begin{equation}
        f(\boldsymbol{x}) = \boldsymbol{W}[\boldsymbol{x}_{eeg},\boldsymbol{x}_{em}] + b,
    \end{equation}
    where the $\boldsymbol{W}$ can be represented by a partitioned matrix $[\boldsymbol{W}_{eeg},\boldsymbol{W}_{em}]$. Then the logits output can be formulated as:
    \begin{equation}
        \begin{split}
            f(\boldsymbol{x}) &= (\boldsymbol{W}_{eeg}\boldsymbol{x}_{eeg} + b_{eeg}) + (\boldsymbol{W}_{em}\boldsymbol{x}_{em} + b_{em}) \\
            & = f(\boldsymbol{x}_{eeg}) + f(\boldsymbol{x}_{em}),
        \end{split}
        \label{concatenation output}
    \end{equation}
    where the $f(\boldsymbol{x}), f(\boldsymbol{x}_{eeg}), f(\boldsymbol{x}_{em}) \in\mathbb{R}^{3}$ are the multi-modal logits and uni-modal logits of EEG and EM modalities respectively. 

    In Eq. (\ref{concatenation output}), the final classification logits can be decomposed into two logits calculated independently using the single modality. Hence, the output $f(\boldsymbol{x})_{k}$ of each sample in groud truth class $k$ $(k=1, 2, 3)$ is the sum of the two single-modality's output logits $f(\boldsymbol{x}_{eeg})_{k}+f(\boldsymbol{x}_{em})_{k}$ in class $k$. The logit output by a single modality for the ground truth class can be interpreted as that modality's contribution to classification. Higher logit values indicate stronger contributions to correct classification, suggesting that such modalities should receive greater weights during fusion.

\subsubsection{Contribution-Guided Reweighting}
    \medskip
    Based on the above analyses, we introduce the modal contribution ratio to provide explicit guidance for reweighting module optimization. We first apply the softmax function to convert the logits within each modality into positive contribution scores. The modal contribution ratio is then calculated as the proportion of each modality's contribution relative to their sum. The process can be formulated as follows:
    \begin{equation}
        c_{eeg} = \sum_{k=1}^{3}\mathbb{I}_{(k=y_{true})}\cdot \sigma \left(\boldsymbol{W}_{eeg}\boldsymbol{x}_{eeg} + b_{eeg}\right)_{k}, 
    \end{equation}
    \begin{equation}
        c_{em} = \sum_{k=1}^{3}\mathbb{I}_{(k=y_{true})}\cdot \sigma \left(\boldsymbol{W}_{em}\boldsymbol{x}_{em} + b_{em}\right)_{k}, 
    \end{equation}
    \begin{equation}
        r_{eeg} = \dfrac{c_{eeg}}{c_{eeg}+c_{em}}, \qquad r_{em} = \dfrac{c_{em}}{c_{eeg}+c_{em}},
    \end{equation}
    where $\sigma(\cdot)$ represents the softmax function and $\mathbb{I}_{(\cdot)}$ is an indicator asserting the subscript condition, which outputs a mask tensor with element 0 or 1. $c_{eeg}$ and $c_{em}$ denote the contribution of EEG and EM modalities to classification respectively. $r_{eeg}$ and $r_{em}$ denote the contribution ratio of EEG and EM modalities respectively.

    For modal reweighting, we use a simple linear layer following the softmax function as a weight prediction network to learn the contribution ratios of different modes as modal weights for concatenation fusion. We use $\phi(\cdot)$ to represent the weight prediction network, and the reweighting process can be represented as follows:
    \begin{equation}
        \boldsymbol{x}_{f} = \left[\boldsymbol{x}_{eeg}\cdot\phi([\boldsymbol{x}_{eeg},\boldsymbol{x}_{em}])_{0},\boldsymbol{x}_{em}\cdot\phi([\boldsymbol{x}_{eeg},\boldsymbol{x}_{em}])_{1}\right],
    \end{equation}
    where the $\phi([\boldsymbol{x}_{eeg},\boldsymbol{x}_{em}]) \in \mathbb{R}^{2}$ is optimized by the contribution-guided loss $\mathcal{L}_{cg}$:
    \begin{equation}
        \mathcal{L}_{cg} = \dfrac{1}{N}\sum_{n=1}^{N}||(r_{eeg}^{n},r_{em}^{n}) - \phi([\boldsymbol{x}_{eeg}^{n},\boldsymbol{x}_{em}^{n}])||_{L1},
        \label{contribution-guided loss}
    \end{equation}
    where $||\cdot||_{L1}$ represents the $L1$-norm and $N$ denotes the total number of samples. $\boldsymbol{x}_{eeg}^{n}$ and $\boldsymbol{x}_{em}^{n}$ represent the flattened features from the DCM for EEG and EM modalities of the n-th sample respectively, while $r_{eeg}^{n}$ and $r_{em}^{n}$ denote their corresponding contribution ratios.  

\subsection{Hierarchical Self-distillation Module}
    In multi-class target RSVP tasks, both target-1 and target-2 classes exhibit significant discriminative characteristics from the non-target class \cite{wei2024preliminary}. Given that many studies have demonstrated robust discrimination between target and non-target classes in RSVP decoding \cite{PPNN,ji2024novel,mao2023cross}, we propose a hierarchical learning framework that decomposes the model into two levels: a foundational binary classification task and a higher-level three-class discrimination task. This hierarchical architecture leverages self-distillation to transfer knowledge from the high-performing binary classifier to guide the more challenging three-class discrimination, thereby reducing misjudgment of non-target instances.

    The input to the hierarchical self-distillation module (HSM) is the fusion feature $\boldsymbol{x}_{f} \in \mathbb{R}^{(2\times c\times d)}$ output by the CG-RM. Our hierarchical framework uses two different linear layers as binary and triplet classifiers, denoted as $f^{bin}(\cdot)$ and $f^{tri}(\cdot)$ respectively. The output of the binary classifier $f^{bin}(\boldsymbol{x}_{f})\in\mathbb{R}^{2}$ has two values denoting the non-target and target classes, respectively. The output of the triplet classifier $f^{tri}(\boldsymbol{x}_{f})\in\mathbb{R}^{3}$ has three values denoting the non-target, target-1, and target-2 classes, respectively. To align the output of the two classifiers, we modify the output of the three-class classifier to a binary classifier $f^{mtri}(\cdot)$:
    \begin{equation}
        f^{mtri}(\boldsymbol{x}_{f}) = \left[\sigma\left(f^{tri}(\boldsymbol{x}_{f})\right)_{0},\sigma\left(f^{tri}(\boldsymbol{x}_{f})\right)_{1}+\sigma\left(f^{tri}(\boldsymbol{x}_{f})\right)_{2}\right],
    \end{equation}
    where the sum of the probabilities of target-1 and target-2 $\sigma\left(f^{tri}(\boldsymbol{x}_{f})\right)_{1}+\sigma\left(f^{tri}(\boldsymbol{x}_{f})\right)_{2}$ represents the probability that the sample belongs to the target class, and $\sigma(\cdot)$ denotes the softmax function. In self-distillation learning, we adopt symmetric Kullback-Leibler divergence $\mathcal{L}_{skl}(\cdot,\cdot)$ \cite{hinton2015distilling} as a measure of self-distillation loss $\mathcal{L}_{sd}$ to maximize the consistency between the prediction probability distributions of two classifiers:
    \begin{equation}
        \mathcal{L}_{sd} = \dfrac{1}{N}\sum_{n=1}^{N}\mathcal{L}_{skl}\left[f^{mtri}(\boldsymbol{x}_{f}^{n}),\sigma\left(f^{bin}(\boldsymbol{x}_{f}^{n})\right)\right].
        \label{self-distillation loss}
    \end{equation}

    Moreover, to fully utilize the binary classifier $f^{bin}(\cdot)$ for reducing the misjudgment of non-target samples, we employ the output of $f^{bin}(\cdot)$ as attention weights for the three-class classifier's outputs. Specifically, we modify $f^{bin}(\cdot)$ to $f^{mbin}(\cdot)$ by duplicating the target class neuron's output to create a third neuron, allowing dimension matching with the three-class logits:
    \begin{equation}
        f^{mbin}(\boldsymbol{x}_{f})=\left[\sigma\left(f^{bin}(\boldsymbol{x}_{f})\right)_{0}, \sigma\left(f^{bin}(\boldsymbol{x}_{f})\right)_{1}, \sigma\left(f^{bin}(\boldsymbol{x}_{f})\right)_{1}\right].
    \end{equation}
    Finally, the predicted probability can be acquired by element-wise multiplication of $f^{mbin}(\boldsymbol{x}_{f})$ and $f^{tri}(\boldsymbol{x}_{f})$, followed by passing through the softmax function. 

\subsection{Loss Function}
    The model optimization with classification loss $\mathcal{L}_{cls}$ consists of three parts. First, we apply cross-entropy loss $\mathcal{L}_{ce}$ to optimize the entire model using the three-class ground truth labels $y_{tri}$ and predictions from the fused features $\boldsymbol{x}_{f}$. Second, we optimize the binary classifier in HSM using binary cross-entropy loss $\mathcal{L}_{bce}$ between binary ground truth labels $y_{bin}$ and binary predictions from $\boldsymbol{x}_{f}$. Finally, to fully optimize each single-modality stream, we introduce intra-modal losses $\mathcal{L}_{intra\mbox{-}eeg}$ and $\mathcal{L}_{intra\mbox{-}em}$, which apply cross-entropy between $y_{tri}$ and the three-class predictions from individual modality features $\boldsymbol{x}_{eeg}$ and $\boldsymbol{x}_{em}$. The complete classification loss $\mathcal{L}_{cls}$ is formulated as:
    \begin{equation}
        \mathcal{L}_{cls} = \mathcal{L}_{ce} + \mathcal{L}_{bce} + \lambda(\mathcal{L}_{intra\mbox{-}eeg} + \mathcal{L}_{intra\mbox{-}em}),
    \end{equation}
    where the $\lambda$ is the coefficient of intra-modal classification loss and we set it to 0.2.
      
    The overall loss function $\mathcal{L}_{overall}$ comprises three components including the classification loss $\mathcal{L}_{cls}$, the contribution-guided loss $\mathcal{L}_{cg}$ (see Eq. (\ref{contribution-guided loss})), and the self-distillation loss $\mathcal{L}_{sd}$ (see Eq. (\ref{self-distillation loss})):
    \begin{equation}
        \mathcal{L}_{overall} = \mathcal{L}_{cls} + \mathcal{L}_{cg} + \mathcal{L}_{sd}.
    \end{equation}

\section{Experiments}
\subsection{Implementation Details and Experimental Setup}
\subsubsection{Implementation Details}
    In the experiment, we utilize all 64 EEG channels, along with EM signals comprising pupil area and gaze coordinates (horizontal X and vertical Y) from both eyes. Both EEG and EM signals are down-sampled at 128 Hz. The MTREE-Net  processes inputs of $(C_{eeg}, T)$ and $(C_{em}, T)$ for EEG and EM signals respectively, where $C_{eeg}=64$ (EEG channels), $C_{em}=6$ (EM components), and $T=128$ (sampling points). The feature extractor outputs tensors of size $(c,d)$, where $c=32$  denotes the number of feature maps and $d=16$ corresponds to the feature dimension $d=\lfloor T/8\rfloor$.
 
    We implement our model within the PyTorch framework. The optimization process employs the Adam optimizer \cite{kingma2014adam} with an initial learning rate of 0.001. The learning rate is dynamically adjusted using a plateau-based scheduler, which reduces the learning rate by 50\% when the validation balanced accuracy shows no improvement for 5 consecutive epochs. The model is trained with a batch size of 128 for a maximum of 120 epochs. To address the inherent class imbalance in RSVP tasks, we apply a class-balanced resampling strategy during training. Specifically, we down-sample the majority class (non-target) to match the sample size of the minority classes (target-1 and target-2) in the training set, while maintaining the original class distribution in the validation and test sets.

\subsubsection{Experimental Setup}
    We conduct three independent multi-class target RSVP experiments in Section III and collect data in each task. These three tasks can be considered as three independent RSVP datasets, and we perform experiments on them separately. The model assessment follows within-subject decoding settings using the cross-validation strategy. For each subject, we employ a 5-fold cross-validation to partition experimental blocks into training and test sets. Each block serves as the test set once, while the remaining blocks form the training set. Within each training fold, a secondary 5-fold cross-validation subdivides the data in training blocks into training and validation sets. The subject-specific performance is computed by averaging the model's performance across all test blocks. The overall model performance for each task is then determined by averaging the decoding performance across all 43 subjects.

\subsection{Comparison Methods}
    We compare our proposed model with eight EEG decoding methods and three EEG and EM fusion methods. 
    
    For fair comparisons, we resample our EEG signals as described in the comparison methods and modify the channel number to the same as ours. If the codes for the method are open-source, we use open-source codes for verification. If the code is not open-source, we re-implement the model architecture of the comparison methods exactly and optimize the models using the same optimization techniques and parameter settings as described in the source literature. The comparison methods are as follows:

    \subsubsection{EEG Decoding Methods}
    \medskip
    \begin{itemize}
        \item \textbf{EEGNet} \cite{EEGNet}: a CNN-based method with depthwise and separable convolution layers. The implementation is based on the code available at https://github.com/vlawhern/arl-eegmodels.
        \item \textbf{MS-CNN} \cite{wang2021performance}: a multi-scale
        convolutional neural network model consisting of seven layers. The implementation follows the description in \cite{wang2021performance}.
        \item \textbf{PLNet} \cite{PLNet}: a CNN-based method that utilizes the phase-locked characteristic of ERP signals to extract spatiotemporal features. The method is implemented according to \cite{PLNet}.
        \item \textbf{PPNN} \cite{PPNN}: a CNN-based method employing dilated temporal convolution layers to preserve phase information. The implementation is based on the description in \cite{PPNN}.
        \item \textbf{EEG-conformer} \cite{song2022eeg}: a compact convolutional Transformer that encapsulates local and global features within a unified EEG classification framework. The implementation is based on the code available at https://github.com/eeyhsong/EEG-Conformer.
        \item \textbf{STSTNet} \cite{luo2023dual}: a dual-branch spatio-temporal-spectral Transformer incorporating a channel attention weighting module. The implementation refers to https://github.com/ljbuaa/VisualDecoding.
        \item \textbf{MDCNet} \cite{wei2024preliminary}: a CNN-based method designed for multi-class target RSVP decoding, which utilizes multi-scale dilated convolutions to extract EEG amplitude and phase information in the temporal dimension. The implementation refers to \cite{wei2024preliminary}.
        \item \textbf{HCANN} \cite{ji2024novel}: a CNN-based method combining depthwise separable convolution and the multi-head mechanism to adaptively extract EEG temporal and spatial representations. We utilize the open-source implementation available at https://github.com/youshuoji/HCANN.
        \item \textbf{EEG baseline}: The EEG baseline model is a component of MTREE-Net, consisting of the EEG feature extractor and a linear layer. This model uses only EEG signals for classification and is optimized using the loss function $\mathcal{L}_{ce}$.
    \end{itemize}

    \subsubsection{EEG and EM Fusion Methods}
    \medskip
    \begin{itemize}
        \item \textbf{mDCAN} \cite{zhao2021multimodal}: a modified variant of DCAN \cite{nguyen2018improved} to capture the inter-modality relations. We reimplement the mDCAN according to \cite{zhao2021multimodal}, based on the implementation of the Dense Co-Attention Network (DCAN) available in the repository at https://github.com/cvlab-tohoku/Dense-CoAttention-Network.
        \item \textbf{CMGFNet} \cite{mao2023cross}: a cross-modal guiding and fusion network that guides EM modality features to complement the EEG modality and enhances the multi-modal features by exploring both intra-modal and inter-modal interactions. The implementation refers to \cite{mao2023cross}.
        \item \textbf{FGFRNet} \cite{fu2024cross}: an EEG and EM fusion network that employs EEG features to guide the feature extraction of EM features and explore the task-related information within EM signals. The implementation refers to \cite{fu2024cross}. 
    \end{itemize}

\subsection{Evaluation Metrics}
    We employ Balanced-Accuracy (BA), recall, and F1-score as evaluation metrics. Balanced accuracy measures the proportion of correctly classified samples among all samples by calculating the arithmetic mean of the accuracy across all three classes. Recall which is computed as the arithmetic mean of the recall values for the two target categories, assesses the model's ability to identify target samples. The F1-score is the arithmetic mean of the F1-score of the three categories, which is the harmonic mean of precision and recall, making it suitable for imbalanced datasets. The calculation of these metrics can be formulated as follows:
    
    \begin{small}
        \begin{equation}
        TPR_{k} = \dfrac{TP_{k}}{TP_{k}+FN_{k}},
        \end{equation}
    \end{small}

    \begin{small}
        \begin{equation}
            F1_{k} = \frac{2\times precision_{k}\times TPR_{k}}{precision_{k}+TPR_{k}},
        \end{equation}
    \end{small}

    \begin{small}
        \begin{equation}
            BA = \frac{1}{3}\sum_{k=1}^{3}TPR_{k}, \enspace Recall = \frac{1}{2}\sum_{k=2}^{3}TPR_{k}, \enspace F1 = \frac{1}{3}\sum_{k=1}^{3}F1_{k},
        \end{equation}
    \end{small}
    where $k=1, 2, 3$ corresponds to the non-target, target-1 and target-2 categories respectively. $TP_{k}$ stands for the number of correctly classified positive samples in the $k$-th class, $FN_{k}$ denotes the number of incorrectly classified positive samples in the $k$-th class, and $FP_{k}$ signifies the number of incorrectly classified negative samples in the $k$-th class. The $precision_{k}$ is a precision metric that measures the proportion of correctly classified positive samples among all samples classified as positive in the $k$-th class.

\subsection{Statistical Analysis}
    We conduct one-way and two-way repeated measures analysis of variance (ANOVA) to evaluate the effects of different factors on RSVP decoding performance. If the data violate the sphericity assumption, the Greenhouse-Geisser correction is applied. For each significant factor, post-hoc analyses are performed and the Holm-Bonferroni correction is used to adjust the p-values in multiple comparisons. All statistical tests used 0.05 as the significance threshold.

\section{Results}

\subsection{Comparison Experiment}
    \setlength{\tabcolsep}{1.5mm}{
    \begin{table*}[ht]
        \begin{threeparttable}
            \small
            \renewcommand\arraystretch{1.3}
            \centering
            \caption{Classification performance of MTREE-Net and different comparison methods on three tasks (mean).}
            \label{comparison experiment}
            \begin{tabular}{lllllllllll}
                \toprule[1.2pt]
                \multirow{3}{*}{\textbf{ Methods}}&\multirow{3}{*}{\textbf{ Modality}}& \multicolumn{3}{c}{\multirow{2}{*}{\textbf{ Task A}}} & \multicolumn{3}{c}{\multirow{2}{*}{\textbf{ Task B}}} & \multicolumn{3}{c}{\multirow{2}{*}{\textbf{ Task C}}}\\
                & & \multicolumn{3}{c}{} & \multicolumn{3}{c}{} & \multicolumn{3}{c}{}\\
                \cmidrule(lr){3-5}\cmidrule(lr){6-8}\cmidrule(lr){9-11}
                & & \textbf{BA $(\%)$}& \textbf{Recall $(\%)$} & \textbf{F1} & \textbf{BA $(\%)$} & \textbf{Recall $(\%)$} & \textbf{F1} & \textbf{BA $(\%)$} & \textbf{Recall $(\%)$} & \textbf{F1} \\
              \hline
                EEGNet& EEG &$56.81^{\star\star\star} $ & $49.14^{\star\star\star} $& $0.3856^{\star\star\star}  $&   $60.13^{\star\star\star} $ & $54.67^{\star\star\star} $& $0.3927^{\star\star\star} $&  $59.53^{\star\star\star}$ & $51.85^{\star\star\star} $& $0.4066^{\star\star\star} $ \\
                MS-CNN&EEG &  $57.94^{\star\star\star} $ & $47.54^{\star\star\star} $& $0.4186^{\star\star\star}  $&   $61.05^{\star\star\star} $ & $52.77^{\star\star\star} $& $0.4300^{\star\star\star} $&  $60.06^{\star\star\star} $ & $49.98^{\star\star\star} $& $0.4362^{\star\star\star} $ \\
                PLNet &EEG &  $56.17^{\star\star\star} $ & $51.10^{\star\star\star} $& $0.3620^{\star\star\star}  $&   $59.30^{\star\star\star} $ & $57.41^{\star\star\star} $& $0.3589^{\star\star\star} $&  $59.03^{\star\star\star} $ & $54.26^{\star\star\star} $& $0.3791^{\star\star\star} $ \\
                PPNN&EEG &  $61.10^{\star\star\star} $ & $52.42^{\star\star\star} $& $0.4275^{\star\star\star} $&   $65.42^{\star\star\star} $ & $59.78^{\star\star\star} $& $ 0.4389^{\star\star\star}$&  $  64.00^{\star\star\star}$ & $ 55.68^{\star\star\star}$& $ 0.4516^{\star\star\star}$ \\
                EEG-conformer&EEG&  $60.02^{\star\star\star} $ & $51.11^{\star\star\star} $& $0.4203^{\star\star\star}  $&   $63.99^{\star\star\star} $ & $57.53^{\star\star\star} $& $0.4329^{\star\star\star} $&  $62.90^{\star\star\star} $ & $54.68^{\star\star\star} $& $ 0.4389^{\star\star\star}$ \\
                STSTNet&EEG&  $59.40^{\star\star\star} $ & $55.44^{\star\star\star} $& $0.3737^{\star\star\star}  $&   $64.27^{\star\star\star} $ & $63.02^{\star\star\star} $& $0.3873^{\star\star\star} $&  $62.58^{\star\star\star} $ & $59.04^{\star\star\star} $& $0.3931^{\star\star\star} $ \\
                MDCNet&EEG &  $63.03^{\star\star\star} $ & $56.66^{\star\star} $& $0.4217^{\star\star\star}  $&   $69.02^{\star\star\star} $ & $\underline{65.63}^{\star\star} $& $0.4468^{\star\star\star} $&  $67.47^{\star\star\star} $ & $\underline{61.63}^{\star\star} $& $0.4570^{\star\star\star} $ \\
                HCANN &EEG&  $60.32^{\star\star\star} $ & $51.07^{\star\star\star} $& $ 0.4284^{\star\star\star} $&   $64.29^{\star\star\star} $ & $58.03^{\star\star\star} $& $0.4383^{\star\star\star} $&  $ 63.25^{\star\star\star}$ & $54.62^{\star\star\star} $& $0.4509^{\star\star\star} $ \\
                \hline
                mDCAN &EEG, EM &  $ 61.94^{\star\star\star}$ & $55.12^{\star\star\star} $& $0.4168^{\star\star\star}  $&   $66.32^{\star\star\star} $ & $62.76^{\star\star\star} $& $0.4253^{\star\star\star} $&  $ 64.68^{\star\star\star}$ & $58.52^{\star\star\star} $& $0.4346^{\star\star\star} $ \\
                CMGFNet &EEG, EM &  $ \underline{65.64}^{\star\star\star}$ & $55.07^{\star\star\star} $& $\underline{0.5000}^{\star\star\star}  $&   $\underline{70.58}^{\star\star\star} $ & $63.19^{\star\star\star} $& $\underline{0.5211}^{\star\star\star} $&  $ \underline{69.01}^{\star\star\star}$ & $59.14^{\star\star\star} $& $\underline{0.5378}^{\star\star\star} $ \\
                FGFRNet &EEG, EM &  $63.44^{\star\star\star} $ & $54.00^{\star\star\star} $& $0.4594^{\star\star\star}  $&   $67.55^{\star\star\star} $ & $61.04^{\star\star\star} $& $0.4738^{\star\star\star} $&  $66.67^{\star\star\star} $ & $57.98^{\star\star\star} $& $0.4888^{\star\star\star} $ \\
                \hline
                EEG baseline &EEG&  $64.22^{\star\star\star} $ & $\underline{56.70}^{\star\star\star} $& $0.4433^{\star\star\star}  $&   $69.33^{\star\star\star} $ & $64.83^{\star\star\star} $& $0.4618^{\star\star\star} $&  $67.52^{\star\star\star} $ & $60.33^{\star\star\star} $& $0.4727^{\star\star\star} $ \\
                %Ours&EEG, EM &  $\mathbf{68.89} $ & $\mathbf{59.31} $& $\mathbf{0.5243}  $&   $\mathbf{74.42} $ & $ \mathbf{68.36}$& $\mathbf{0.5502} $&  $\mathbf{73.00} $ & $\mathbf{64.16} $& $\mathbf{0.5776} $ \\
                MTREE-Net&EEG, EM &  $\mathbf{68.96} $ & $\mathbf{59.21} $& $\mathbf{0.5280}  $&   $\mathbf{74.42} $ & $ \mathbf{68.15}$& $\mathbf{0.5537} $&  $\mathbf{73.05} $ & $\mathbf{64.12} $& $\mathbf{0.5799} $ \\
             \bottomrule[1.2pt] 
            \end{tabular}
            \begin{tablenotes}
                \item \footnotesize The asterisks in the table indicate a significant difference between the performance of MTREE-Net and the comparison methods by paired t-tests ($\star p<0.05,\star\star p<0.01,\star\star\star p<0.001$).
            \end{tablenotes}
        \end{threeparttable}
    \end{table*}}
    
    Table \ref{comparison experiment} presents the comparative results between MTREE-Net and different comparison methods on three tasks. A two-way repeated measures ANOVA on BA, recall, and F1-score reveals significant main effects for both methods (all metrics: $p<0.001$) and tasks (all metrics: $p<0.001$), along with significant interaction effects between these two factors (all metrics: $p<0.01$). In each task, post-hoc tests demonstrate that our proposed MTREE-Net achieves significantly higher BA compared to both EEG decoding and EEG-EM fusion methods in all three tasks (all: $p<0.001$). Specifically, MTREE-Net also exhibits superior performance in terms of recall (all: $p<0.01$) and F1-score (all: $p<0.001$) in each task. These results demonstrate the effectiveness of our method in enhancing multi-class target RSVP decoding performance.
    
    Among EEG decoding methods, the MDCNet which is specifically designed for multi-class RSVP decoding, demonstrates significantly higher BA (all: $p<0.05$) and recall (all: $p<0.05$) in all three tasks. It also achieves the best F1-scores in Task B and C. Our EEG baseline significantly outperforms MDCNet in BA on Task A ($p<0.001$) and in F1-scores on all three tasks (all: $p<0.001$). It also shows comparable performance in BA on Task B and C (both: $p=1.00$), as well as in recall on Task A and B (both: $p>0.35$). These results indicate the effectiveness of our feature extractor based on multi-scale convolutions for RSVP-EEG feature extraction. Furthermore, MTREE-Net demonstrates significantly superior performance compared to the EEG baseline across three tasks (all: $p<0.001$), which demonstrates the efficacy of EEG and EM fusion for enhancing decoding performance. In multi-modal fusion methods, CMGFNet achieves the highest BA (all: $p<0.001$) and F1-scores (all: $p<0.001$) among comparison methods in all tasks and the MTREE-Net significantly outperforms the CMGFNet in terms of BA, recall, and F1-score in all tasks (all: $p<0.001$), which reveals the superiority of our proposed fusion strategies in EEG and EM feature integration compared to existing multi-modal approaches.

\subsection{Ablation Study}
    We conduct ablation studies to evaluate the efficacy of our proposed dual-complementary module (DCM), contribution-guided reweighting module (CG-RM), and hierarchical self-distillation module (HSM) in MTREE-Net, respectively. Firstly, we evaluate the performance of MTREE-Net without DCM (w/o DCM), where the outputs of feature extractors are directly fed to CG-RM for fusion. Then, we conduct two experiments for the CG-RM ablation: (1) removing the entire CG-RM module (w/o CG-RM) where features are concatenated directly, and (2) removing only the contribution-guided loss (w/o $\mathcal{L}_{cg}$) while employing the reweighting fusion without contribution guidance. Finally, we similarly perform two experiments for the HSM ablation: (1) removing the entire HSM module (w/o HSM) while only leaving the triplet classifier for classification, and (2) removing only the self-distillation loss (w/o $\mathcal{L}_{sd}$) while preserving the hierarchical structure. All the ablation experiments are performed on three tasks.

    \setlength{\tabcolsep}{1.6mm}{
    \begin{table*}[htbp]
        \begin{threeparttable}
            \small
            \centering
            \renewcommand\arraystretch{1.3}
            \caption{Ablation study of MTREE-Net on three tasks (mean).}
            \label{ablation study}
            \begin{tabular}{llllllllllllll}
                \toprule[1.2pt]
                \multicolumn{1}{c}{\multirow{3}{*}{\textbf{ \makecell{Ablation \\ modules}}}}&\multicolumn{1}{c}{\multirow{3}{*}{\textbf{ Models}}}& \multicolumn{3}{c}{\multirow{2}{*}{\textbf{ Task A}}} & \multicolumn{3}{c}{\multirow{2}{*}{\textbf{ Task B}}} & \multicolumn{3}{c}{\multirow{2}{*}{\textbf{ Task C}}}\\
                & & \multicolumn{3}{c}{} & \multicolumn{3}{c}{} & \multicolumn{3}{c}{}\\
                \cmidrule(lr){3-5}\cmidrule(lr){6-8}\cmidrule(lr){9-11}
                 & & \textbf{BA $(\%)$}& \textbf{Recall $(\%)$} & \textbf{F1} & \textbf{BA $(\%)$} & \textbf{Recall $(\%)$} & \textbf{F1} & \textbf{BA $(\%)$} & \textbf{Recall $(\%)$} & \textbf{F1} \\
              \midrule
             \textbf{DCM} &(w/o) DCM & $67.78^{\star\star\star} $& $58.15^{\star\star\star}  $ & $0.5109^{\star\star\star} $& $73.40^{\star\star\star}  $& $67.28^{\star\star\star} $& $0.5361^{\star\star\star}  $& $ 71.91^{\star\star\star}$ & $63.04^{\star\star\star} $& $0.5596^{\star\star\star}  $ \\
                \midrule
             \multirow{2}{*}{\textbf{CG-RM}} &(w/o) CG-RM & $ 67.88^{\star\star\star}$& $58.34^{\star\star\star}  $ & $0.5093^{\star\star\star} $& $73.29^{\star\star\star}  $& $67.19^{\star\star\star} $& $0.5306^{\star\star\star}  $& $71.57^{\star\star\star} $ & $62.68^{\star\star\star} $& $0.5532^{\star\star\star}  $ \\
                &(w/o) $\mathcal{L}_{cg}$ & $67.66^{\star\star\star} $& $57.37^{\star\star\star}  $ & $0.5191^{\star\star\star} $& $73.17^{\star\star\star}  $& $66.47^{\star\star\star} $& $ 0.5428^{\star\star\star} $& $71.84^{\star\star\star} $ & $62.38^{\star\star\star} $& $ 0.5702^{\star\star} $ \\
                 \midrule
              \multirow{2}{*}{\textbf{HSM}}&(w/o) HSM & $ 67.62^{\star\star\star}$& $57.74^{\star\star\star}  $ & $0.5126^{\star\star\star} $& $72.79^{\star\star\star}  $& $ 66.38^{\star\star\star}$& $  0.5330^{\star\star\star}$& $ 71.02^{\star\star\star}$ & $61.66^{\star\star\star} $& $0.5558^{\star\star\star}  $ \\
                 &(w/o) $\mathcal{L}_{sd}$ & $67.99^{\star\star\star} $& $58.31^{\star\star\star}  $ & $0.5149^{\star\star\star} $& $73.47^{\star\star\star}  $& $67.03^{\star\star\star} $& $0.5436^{\star\star\star}  $& $71.97^{\star\star\star} $ & $62.81^{\star\star\star} $& $0.5683^{\star\star\star}  $ \\
                 \midrule
               \multicolumn{2}{c}{MTREE-Net} &  $\mathbf{68.96} $ & $\mathbf{59.21} $& $\mathbf{0.5280}  $&   $\mathbf{74.42} $ & $ \mathbf{68.15}$& $\mathbf{0.5537} $&  $\mathbf{73.05} $ & $\mathbf{64.12} $& $\mathbf{0.5799} $ \\
              \bottomrule[1.2pt]
            \end{tabular}
            \begin{tablenotes}
                \item \footnotesize The asterisks in the table indicate a significant difference between the performance of MTREE-Net and the ablation methods by paired t-tests ($\star p<0.05,\star\star p<0.01,\star\star\star p<0.001$).
            \end{tablenotes}
        \end{threeparttable}
    \end{table*}}

    Table \ref{ablation study} presents the results of ablation experiments. The two-way repeated measures ANOVA on BA, recall, and F1-score reveal significant main effects for both ablation models (all metrics: $p<0.001$) and tasks (all metrics: $p<0.001$). The analysis also shows a significant interaction effect between these factors on F1-score ($p<0.01$), while no significant interactions are found for BA and recall (both $p>0.14$). Post-hoc tests demonstrate that the MTREE-Net significantly outperforms all ablation models in all three tasks in terms of BA, recall, and F1-score (all: $p<0.01$). For DCM ablation, removing the DCM (w/o DCM) leads to significant performance degradation compared to MTREE-Net on all three tasks in terms of BA, recall, and F1-score (all: $p<0.001$), Which validates the effectiveness of DCM in improving decoding performance by realizing feature complementarity across two modalities. Similarly, ablating the CG-RM (w/o CG-RM) results in significant performance drops (all metrics: $p<0.001$) compared to MTREE-Net, indicating its importance in effective multi-modal fusion. Furthermore, CG-RM ablation shows that using only the reweighting module without contribution guidance (w/o $\mathcal{L}_{cg}$) significantly decreases both BA in Task A and B and recall in all three tasks compared to the model without CG-RM (w/o CG-RM). This decline suggests that a reweighting module alone cannot effectively improve fusion performance, demonstrating that incorporating modality contribution ratios to guide reweighting module optimization is essential. Simultaneously, MTREE-Net significantly outperforms (w/o $\mathcal{L}_{cg}$) on all three tasks (all: $p<0.01$), which proves the effectiveness of our proposed contribution-guided approach. For HSM ablation, the model retaining only the hierarchical structure without self-distillation (w/o $\mathcal{L}_{sd}$) performs significantly better than completely removing HSM (w/o HSM) (all: $p<0.001$), demonstrating the effectiveness of the hierarchical design. Additionally, MTREE-Net's superior performance over (w/o $\mathcal{L}_{sd}$) proves that self-distillation learning between binary and triplet classifiers is effective. These ablation results collectively demonstrate that each proposed component in MTREE-Net contributes significantly to MTREE-Net's superior performance in multi-class RSVP decoding.

\subsection{Effect of EM Modality}
    To more thoroughly evaluate the effectiveness of integrating the EM modality into multi-class target RSVP decoding, we compare the performance of MTREE-Net with that of the EEG baseline which lacks the EM modality using confusion matrices.

    \begin{figure}[ht]
        \centering
        \subfigure[Task A]{
			\includegraphics[width=0.99\linewidth]{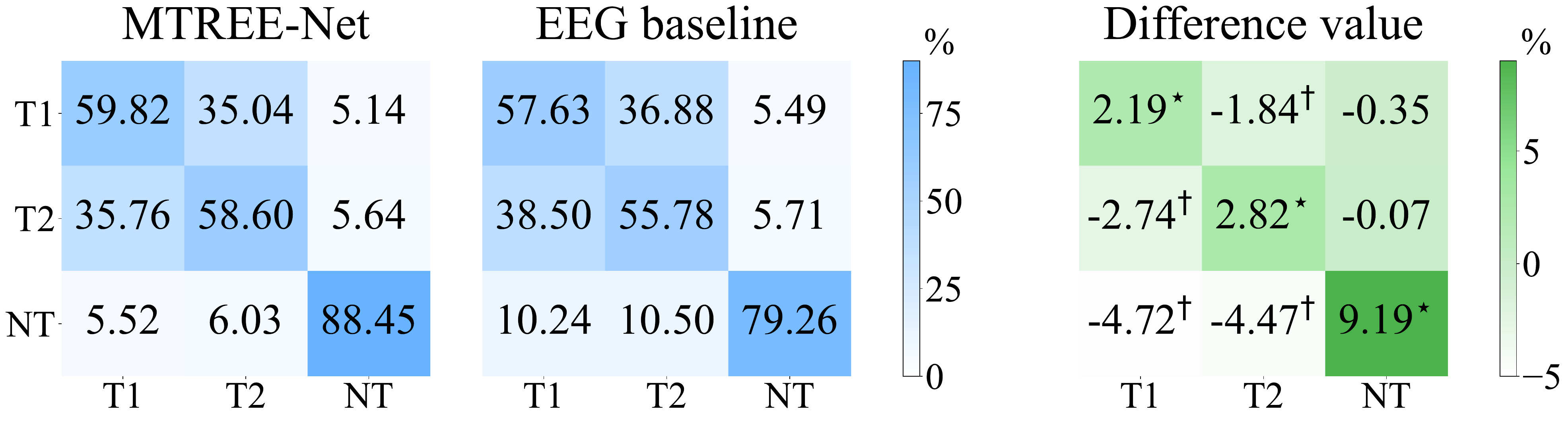}
		}
		\subfigure[Task B]{
			\includegraphics[width=0.99\linewidth]{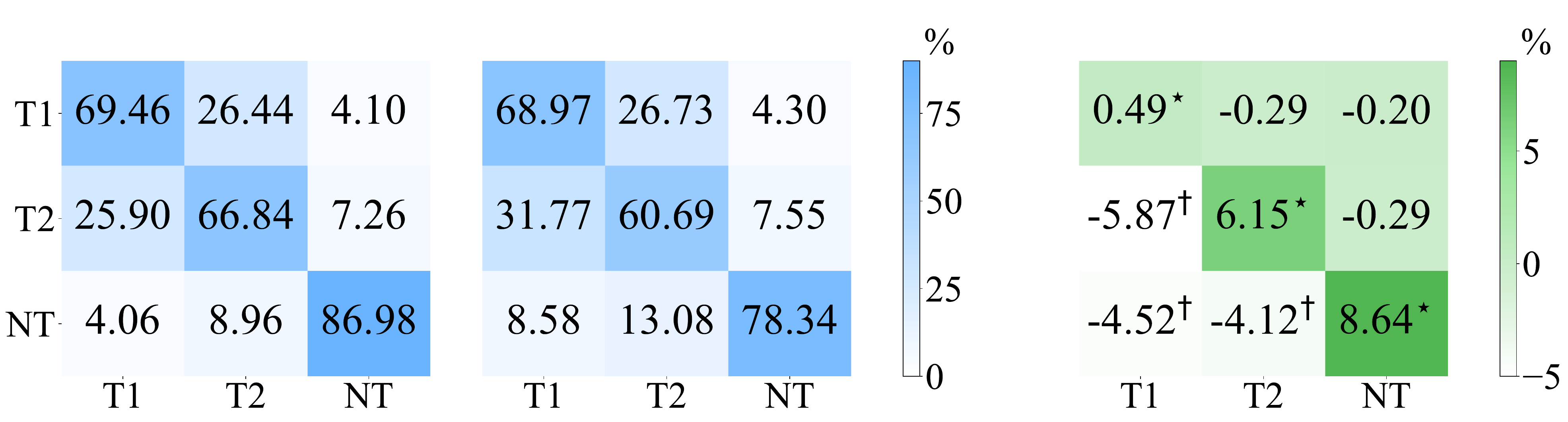}
		}
        \subfigure[Task C]{
			\includegraphics[width=0.99\linewidth]{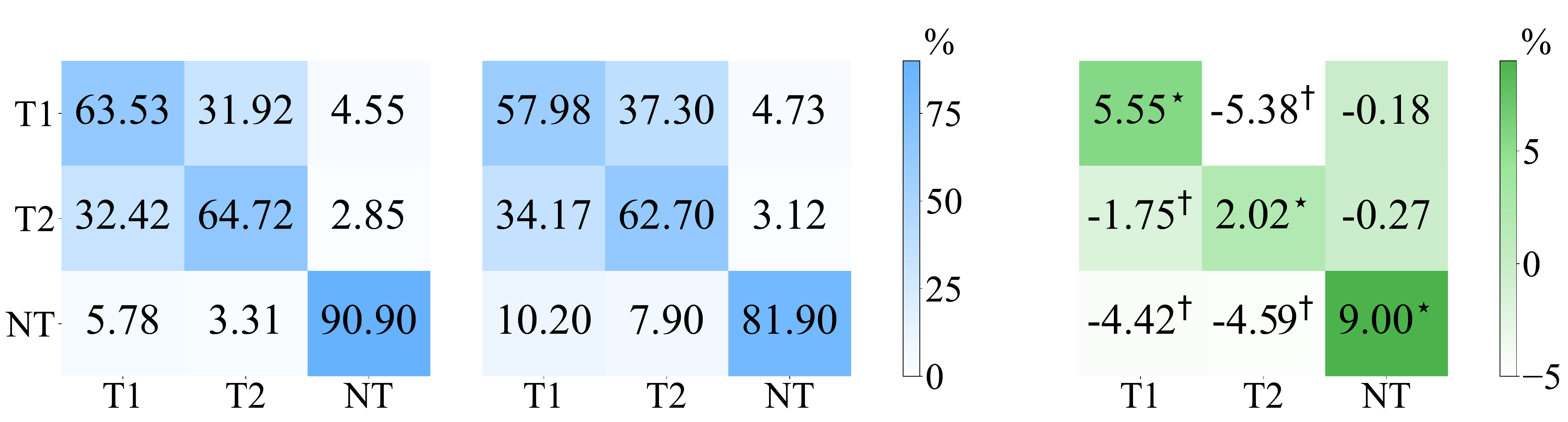}
		}
		\caption{The confusion matrices of MTREE-Net and the EEG baseline, as well as the differences between the results of MTREE-Net and the EEG baseline for (a) Task A, (b) Task B, and (c) Task C. Each confusion matrix displays all metrics as percentages ($\%$), normalized across rows. The labels T1, T2, and NT denote target-1, target-2, and non-target, respectively. The ${\star}$ denotes that the performance metric of MTREE-Net is significantly higher than that of the EEG baseline. At the same time, the ${\dagger}$ indicates that the performance metric of MTREE-Net is significantly lower than that of the EEG baseline ($p<0.05$).}
        \label{confusion matrix}
    \end{figure}

    As shown in Fig. \ref{confusion matrix}, the comparison reveals two key findings: (1) MTREE-Net achieves a significantly higher true positive rate (TPR) than the EEG baseline for two target classes (target-1 and target-2) and non-target class (NT) across all three tasks (all: $p<0.05$). (2) The misclassification rate between target-1 and target-2 is significantly lower for MTREE-Net compared to the EEG baseline across all tasks (all: $p<0.05$), except for Task B, where the misclassification rate of target-1 as target-2 for MTREE-Net tends to be significantly lower than that for EEG baseline ($p=0.18$). (3) The misclassification rate of non-target as either of the two target classes is significantly lower for MTREE-Net compared to the EEG baseline across all tasks (all: $p<0.05$). These results suggest that the integration of the EM modality enhances classification performance by increasing the true positive rate for all three classes and improving the separability between the two target classes, which contributes to mitigating the primary challenge in multi-class RSVP decoding.

\subsection{Effect of Dual-Complementary Module}
    The results in Table \ref{ablation study} demonstrate that our proposed dual-complementary module (DCM) significantly improves the model's decoding performance. To further illustrate the impact of DCM on multi-modal performance, we compare the classification performance between the complete MTREE-Net and the ablation model without DCM (w/o DCM) using multi-modal fusion features and uni-modal features in the multi-modal model, respectively. Table \ref{DCM} presents the BA of MTREE-Net and the ablation model (w/o DCM) using both multi-modal and uni-modal features in three tasks. In the table, ``Fusion" represents the multi-modal decoding performance of the two models, while ``EEG" and ``EM" correspond to the BA achieved using $\boldsymbol{x}_{eeg}$ and $\boldsymbol{x}_{em}$ output by the DCM in MTREE-Net, respectively. For the (w/o DCM) model, since no DCM is used, the uni-modal BA is computed using the flattened features output by the EEG and EM feature extractors in (w/o DCM). The BA calculated using different features for each model are derived from the same trained model, ensuring consistency in the evaluation. 

    \setlength{\tabcolsep}{2.5mm}{
    \begin{table*}[htbp]
        \begin{threeparttable}
            \small
            \renewcommand\arraystretch{1.3}
            \centering
            \caption{Balanced-accuracy ($\%$) of fusion modal and uni-modal (EEG and EM) features in MTREE-Net (mean).}
            \label{DCM}
            \begin{tabular}{cccclllllllll}
                \toprule[1.2pt]
                \multicolumn{4}{c}{\multirow{3}{*}{\textbf{ Models}}}& \multicolumn{3}{c}{\multirow{2}{*}{\textbf{ Task A}}} & \multicolumn{3}{c}{\multirow{2}{*}{\textbf{ Task B}}} & \multicolumn{3}{c}{\multirow{2}{*}{\textbf{ Task C}}}\\
                \multicolumn{4}{c}{}& \multicolumn{3}{c}{} & \multicolumn{3}{c}{} & \multicolumn{3}{c}{}\\
                \cmidrule(lr){5-7}\cmidrule(lr){8-10}\cmidrule(lr){11-13}
                \multicolumn{4}{c}{} & \textbf{Fusion}& \textbf{EEG} & \textbf{EM} & \textbf{Fusion} & \textbf{EEG} & \textbf{EM} & \textbf{Fusion} & \textbf{EEG} & \textbf{EM} \\
              \hline
                \multicolumn{4}{l}{w/o DSM} & $67.78^{\star\star\star} $& $  63.95^{\star\star\star}$ & $38.18^{\star\star\star} $& $73.40^{\star\star\star}  $& $69.57^{\star\star\star} $& $ 37.48^{\star\star\star}$& $71.91^{\star\star\star} $ & $67.92^{\star\star\star} $& $ 36.39^{\star\star\star}$ \\
                \multicolumn{4}{l}{w/ DSM (MTREE-Net)} &  $\mathbf{68.96} $ & $\mathbf{66.39} $& $\mathbf{46.58}  $&   $\mathbf{74.42} $ & $ \mathbf{72.12}$& $\mathbf{45.53} $&  $\mathbf{73.05} $ & $\mathbf{70.92} $& $\mathbf{46.64} $ \\
              \bottomrule[1.2pt]
            \end{tabular}
            \begin{tablenotes}
                \item \footnotesize The asterisks in the table indicate a significant difference between the performance of MTREE-Net and the ablation method by paired t-tests ($\star p<0.05,\star\star p<0.01,\star\star\star p<0.001$).
            \end{tablenotes}
        \end{threeparttable}
    \end{table*}}

    The two-way repeated measures ANOVA on the BA calculated using different features reveal significant main effects for both methods and tasks (all metrics: $p<0.001$), with no significant interaction effects between these factors on BA of fusion features (all metrics: $p=0.93$) and significant interaction effects between these factors on BA of EEG and EM features (both: $p<0.01$). Post-hoc tests show that the BA of MTREE-Net using multi-modal and uni-modal features is significantly higher than that of the (w/o DCM) model in all three tasks (all: $p<0.001$). In particular, without DCM, the highest BA of EM features among the three tasks is only 38.18\%, while the random baseline accuracy for the three-class task is 33.3\%, indicating that the EM-related parameters in the (w/o DCM) model are not optimized effectively and imbalanced optimization occurs during training. After adding DCM, the BA of EM features improved by an average of 8.9\% among the three tasks, indicating that DCM enhances the differentiation of EM features across categories through feature complementarity. Therefore, DCM simultaneously improves the feature differentiation of both EEG and EM across different categories through feature complementarity, thereby enhancing the multi-modal decoding performance of MTREE-Net.

\subsection{Effect of Contribution-Guided Loss}
    To verify that our proposed contribution-guided loss $\mathcal{L}_{cg}$ can improve the performance of multi-modal fusion, we compare the improved performance of MTREE-Net and the ablation model without contribution-guided loss optimization (w/o $\mathcal{L}_{cg}$) using multi-modal features compared to single-modal features. For each model, the improved performance is calculated by measuring the difference between its fusion performance and the best single-modal performance using either EEG ($\boldsymbol{x}_{eeg}$) or EM ($\boldsymbol{x}_{em}$) features output by DCM in the model. We calculate the improved performance of each subject in the three tasks.

    \begin{figure}[!htbp]
        \centering
        \subfigure[Task A]{
			\includegraphics[width=0.32\linewidth]{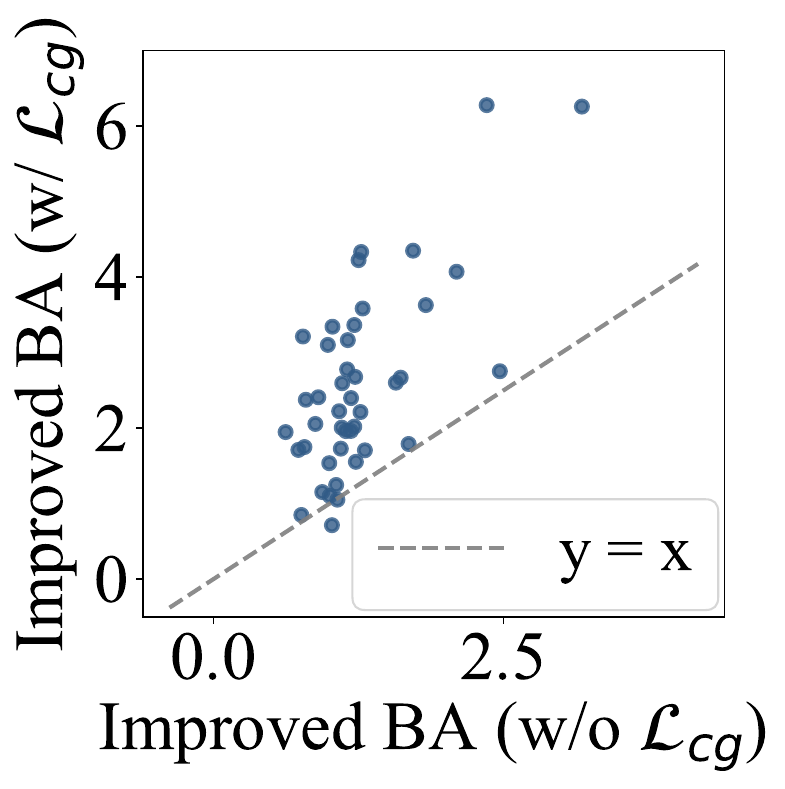}
		}\hspace{-0.3cm}
		\subfigure[Task B]{
			\includegraphics[width=0.32\linewidth]{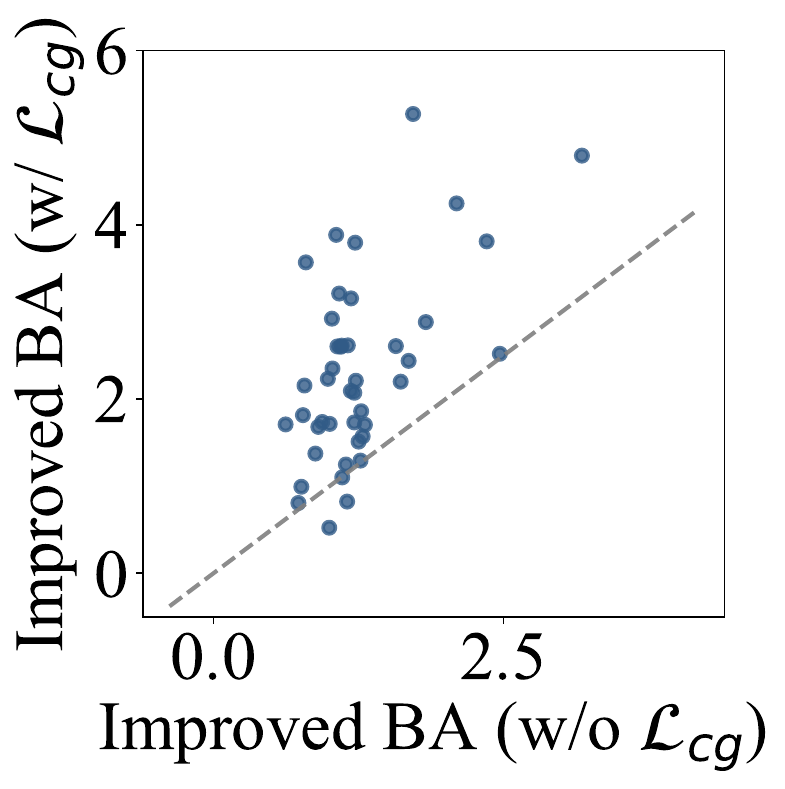}
		}\hspace{-0.3cm}
        \subfigure[Task C]{
			\includegraphics[width=0.32\linewidth]{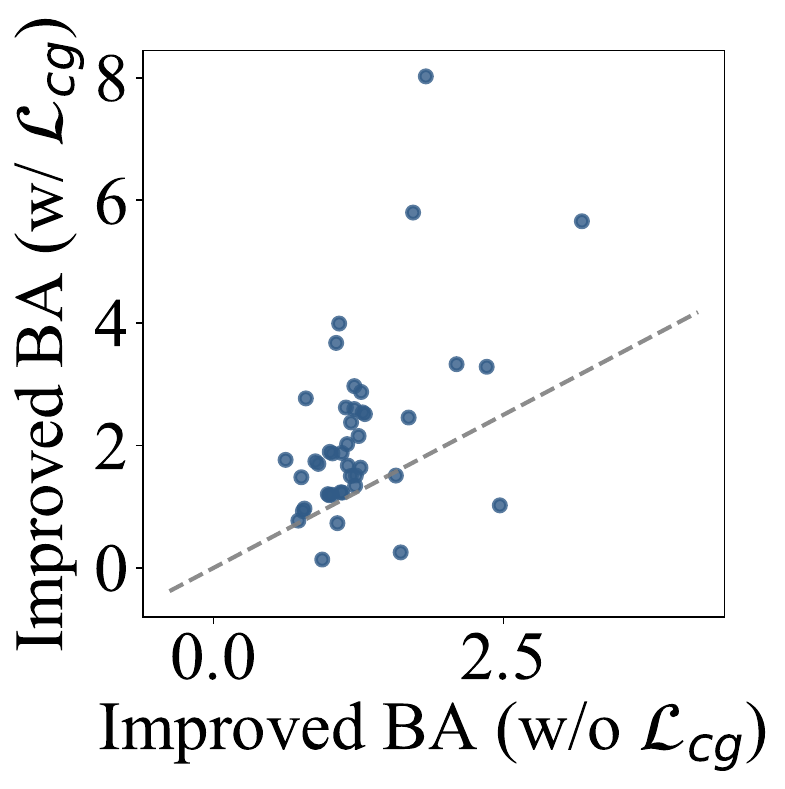}
		}
		\caption{The improvement in BA achieved by MTREE-Net and the ablation model (w/o $\mathcal{L}_{cg}$), compared to the optimal uni-modal BA achieved using the uni-modal features ($\boldsymbol{x}_{eeg}$ or $\boldsymbol{x}_{em}$) within the multi-modal model, is shown for (a) Task A, (b) Task B, and (c) Task C. Each scatter represents the averaged performance of a subject in three tasks, with the x-coordinate indicating the ablation model’s improved BA on each subject, and the y-coordinate showing the BA of complete MTREE-Net (w/ $\mathcal{L}_{cg}$). Scatters above the $y=x$ indicate the improved performance of the model with $\mathcal{L}_{cg}$ on the subject is better than the improved performance without $\mathcal{L}_{cg}$.}
        \label{contribution-guided loss improvements}
    \end{figure}

    Fig. \ref{contribution-guided loss improvements} presents the experimental results, where each point represents a subject with the vertical axis denoting the improved BA of MTREE-Net (w/ $\mathcal{L}_{cg}$) and the horizontal axis indicating the improved BA of (w/o $\mathcal{L}_{cg}$). The results suggest that there are 41, 40, and 38 scatter points are above the $y=x$ in Task A, B, and C, respectively, and the BA improvement with $\mathcal{L}_{cg}$ is significantly higher in all tasks (all: $p<0.001$). This enhancement stems from the contribution-guided loss providing explicit supervisory information to the reweighting module, enabling the module to predict the contribution ratio of each modality and dynamically assign greater weight to the modalities that contribute more to the correct classification. Consequently, MTREE-Net achieves superior fusion performance beyond optimal uni-modal features, which significantly improves decoding performance.
    
\subsection{Effect of Hierarchical Self-distillation Module}
    We present feature visualizations to illustrate the effect of HSM. The t-distributed Stochastic Neighbor Embedding (t-SNE) \cite{van2008visualizing} is employed to project features output by MTREE-Net and the ablation model without HSM (w/o HSM) into a two-dimensional space (see Fig. \ref{tSNE}). For MTREE-Net, we visualize the features output by HSM, and for (w/o HSM), we visualize the features output by CG-RM which are used directly for classification. We perform this analysis on the first three subjects in Task A. For each subject, we train both models using data from the first three blocks and then test them on the remaining blocks. To ensure balanced visualization, we down-sample the non-target class to match the sample sizes of target-1 and target-2 classes.

    \begin{figure}[!htbp]
		\centering
		\subfigure{
			\includegraphics[width=0.99\linewidth]{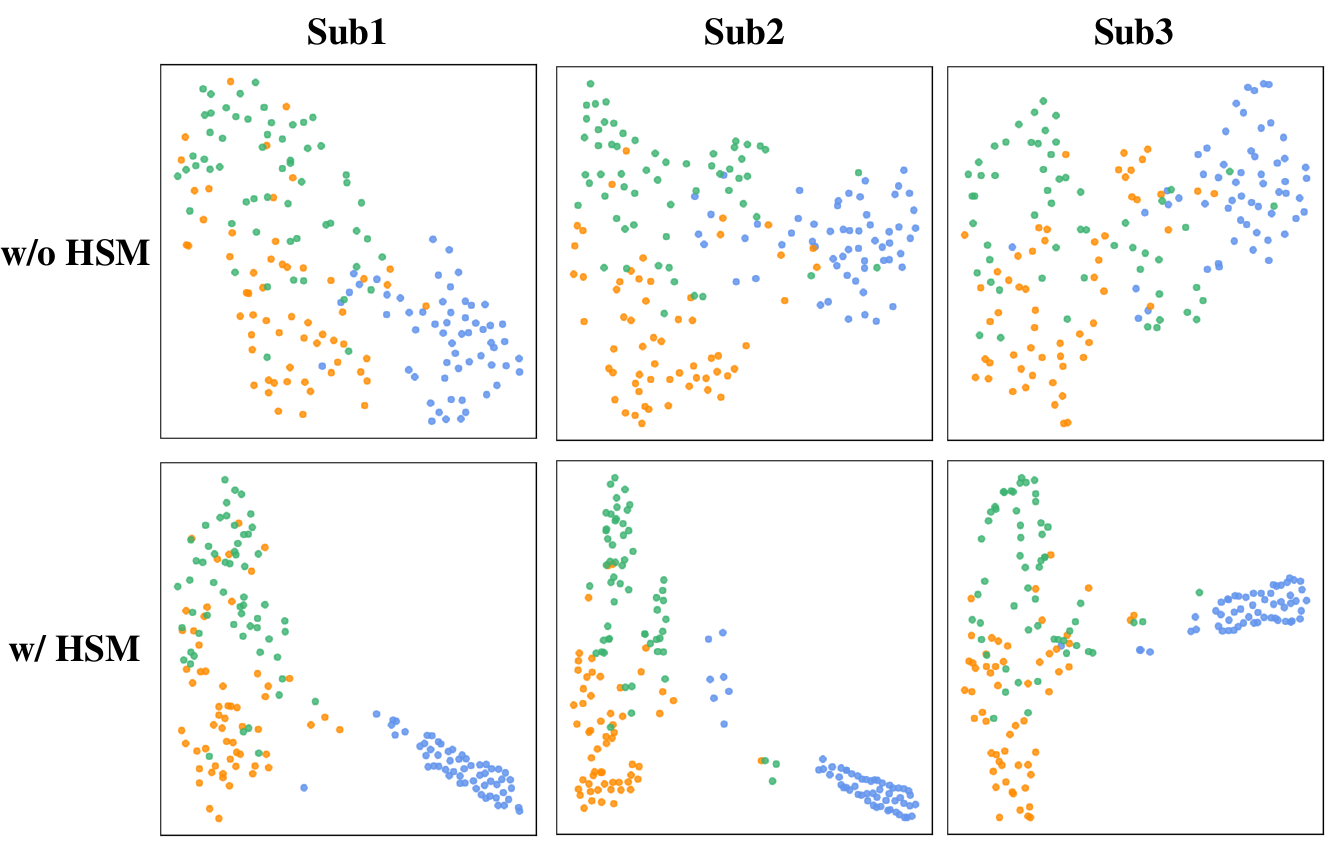}} \\
        \qquad \quad
        \subfigure{
			\includegraphics[width=0.75\linewidth]{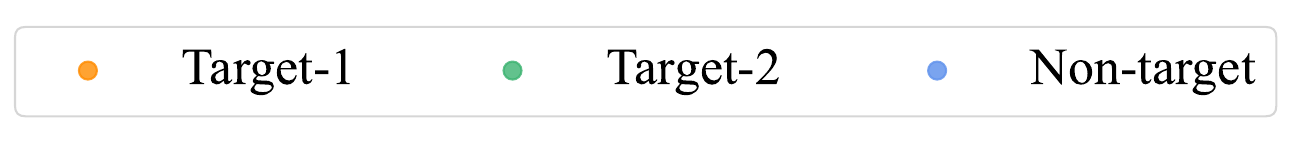}
		}
		\caption{The comparison of the t-SNE visualization results between the ablation model without HSM and MTREE-Net with HSM. The top row displays the ablation model results, while the bottom row shows MTREE-Net results. The three columns represent output features of data from subjects 1, 2, and 3 in Task A respectively. The red, blue, and orange dots indicate the non-target, target-1, and target-2 samples respectively.}
        \label{tSNE}
    \end{figure}

    As illustrated in Figure \ref{tSNE}, for each subject, the output features of (w/o HSM) tend to form tight clusters within each category. However, there is an obvious overlap between the target (target-1 and target-2) and non-target categories, which can lead to the misclassification of non-target instances into target categories. With the integration of HSM, the feature overlap between target and non-target categories in MTREE-Net is significantly reduced and the features of target and non-target classes are linearly separable. This improvement stems from the hierarchical structure where the self-distillation learning transfers knowledge from the high-performing binary classifier to guide the more challenging three-class discrimination, which effectively reduces non-target misclassification rates while preserving discriminative ability among target categories. Consequently, the HSM significantly improves the decoding performance of the model.

    \begin{figure*}[ht]
        \centering
        \subfigure[]{
			\includegraphics[width=0.43\linewidth]{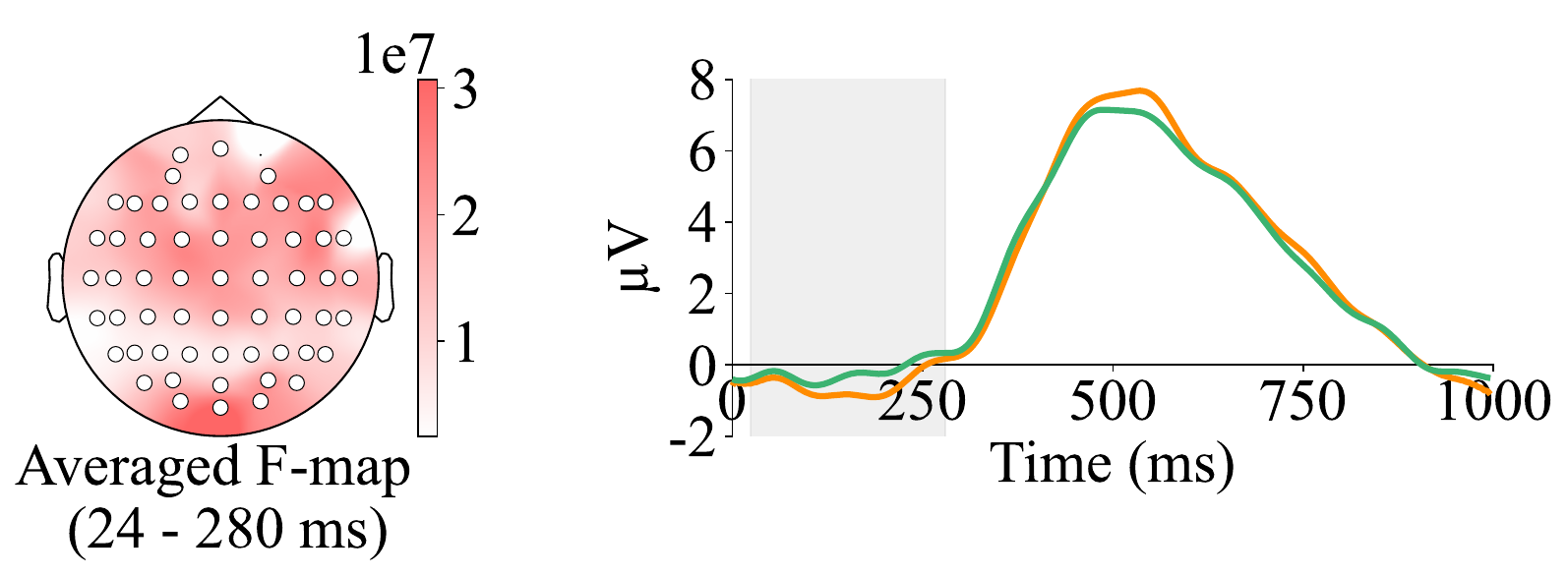}
		}\qquad
		\subfigure[]{
			\includegraphics[width=0.43\linewidth]{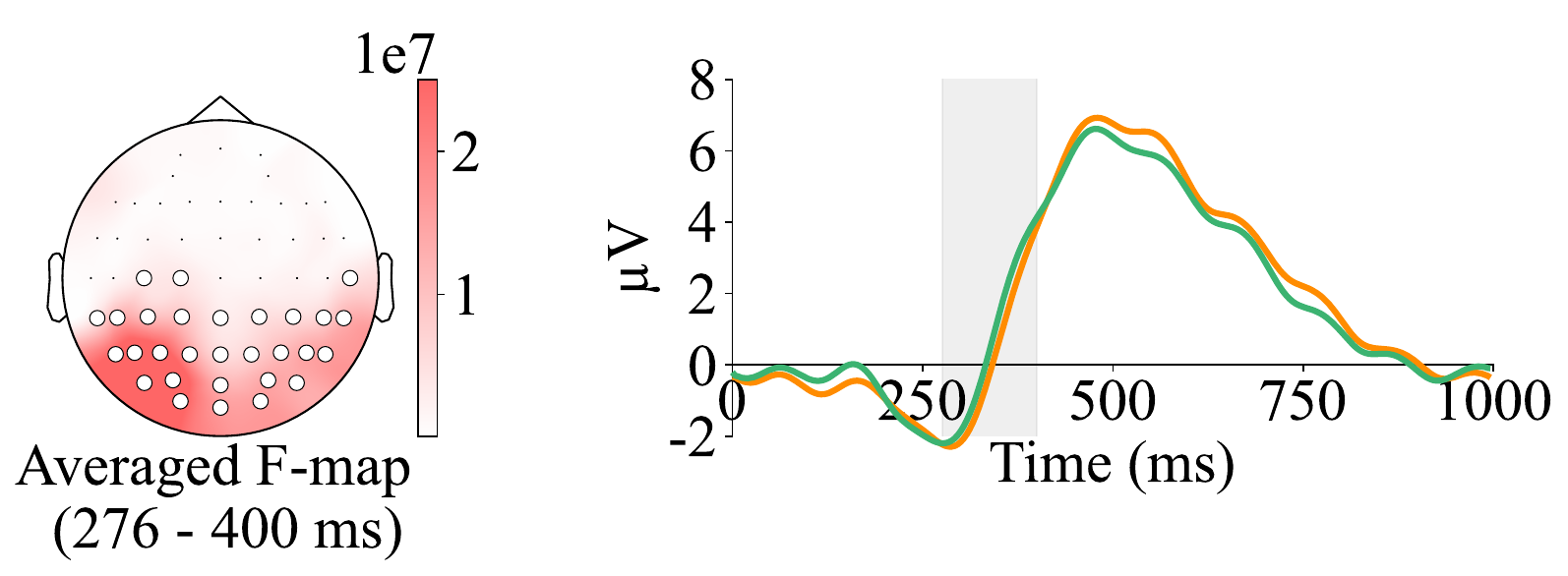}
		}\vspace{-0.2cm}
        \subfigure[]{
			\includegraphics[width=0.43\linewidth]{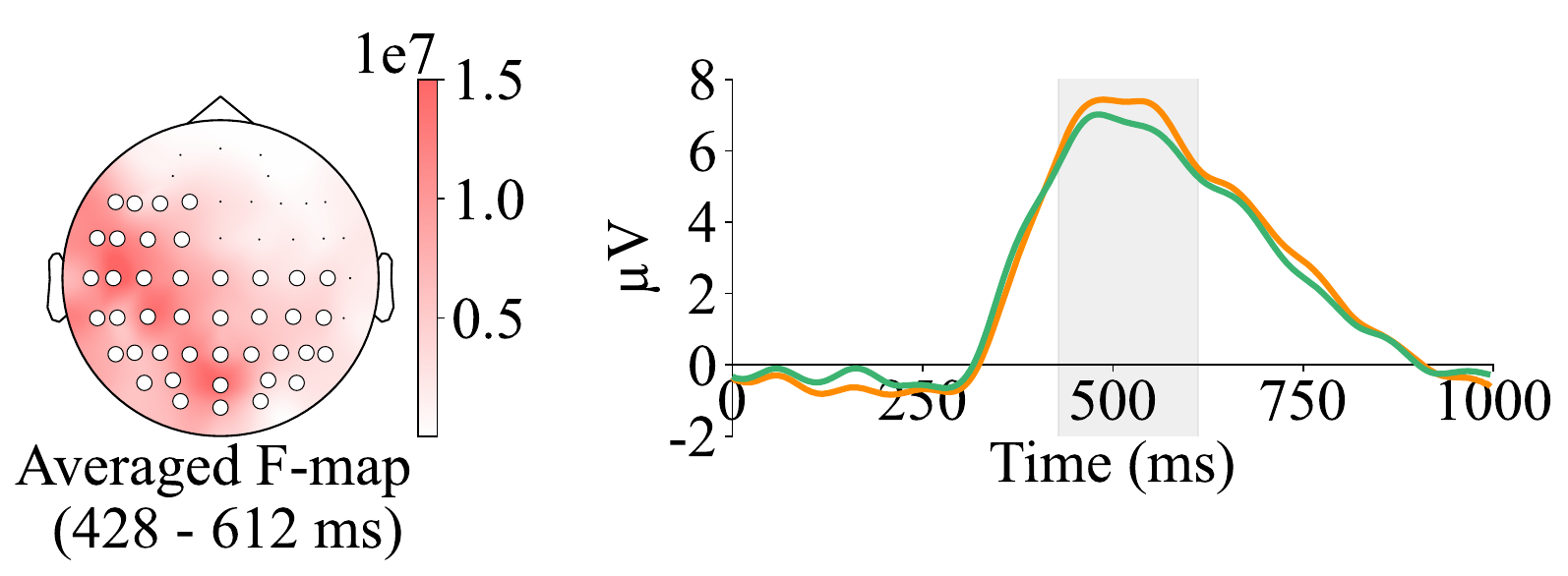}
		}\qquad
        \subfigure[]{
			\includegraphics[width=0.43\linewidth]{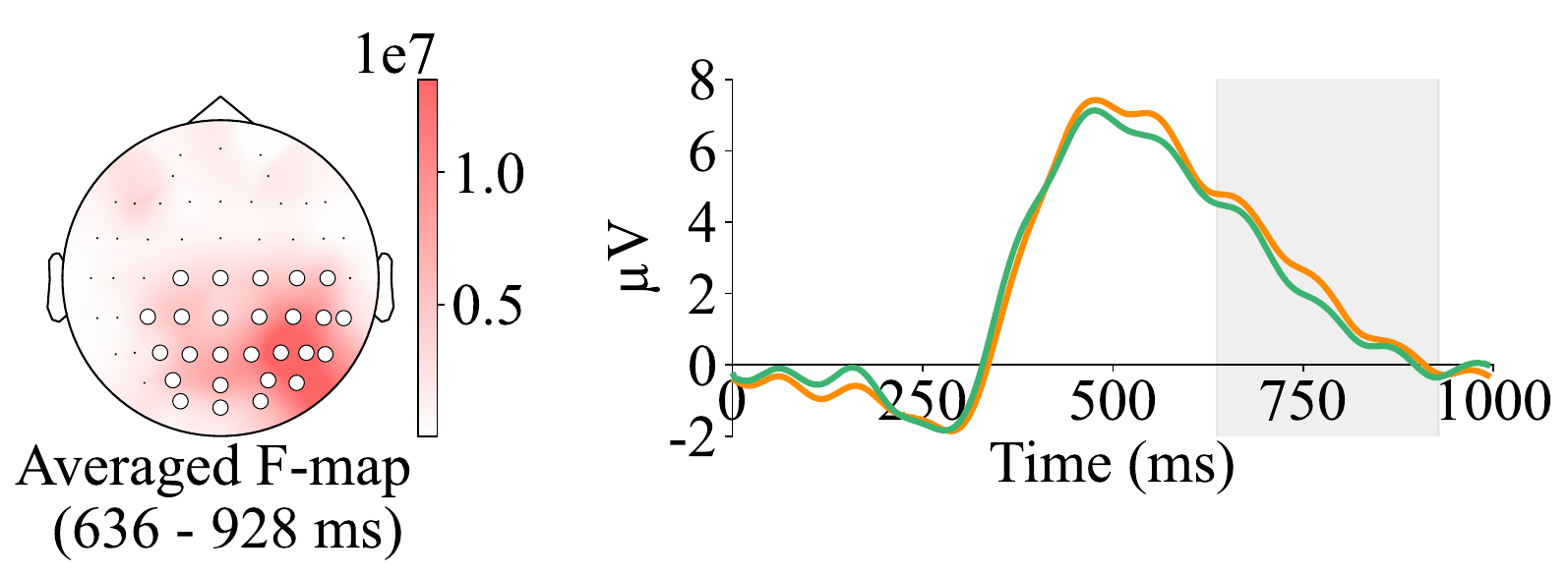}
		}\vspace{-0.06cm}
        \subfigure{
			\includegraphics[width=0.6\linewidth]{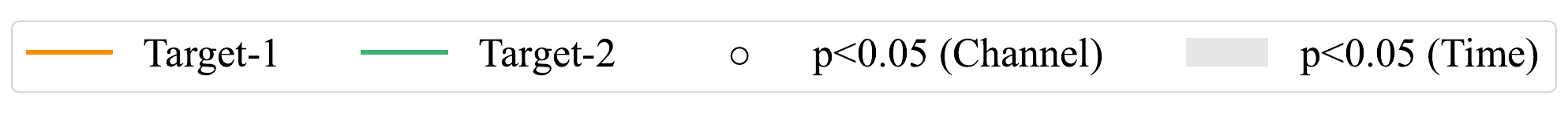}
		}\vspace{-0.1cm}
		\caption{The significant spatiotemporal clusters ($p<0.05$) between target-1 and target-2 conditions non-parametric spatiotemporal cluster test with 1000 permutations (test: one-way repeated measures ANOVA) when comparing both conditions. Each subfigure represents a cluster: the left shows the F-statistics of a significant spatial cluster indicated by white circles over the electrodes, while the right displays the mean activity across channels within the spatial cluster. The light orange shaded regions represent significant ($p<0.05$) temporal clusters.}
        \label{EEG difference}
    \end{figure*}

\subsection{EEG and EM Signals Analysis}
    We employ statistical test methods to analyze EEG and EM signals collected in the multi-class target RSVP task. Since numerous studies have demonstrated significant differences between target and non-target EEG and EM responses in RSVP tasks \cite{mao2023cross,luo2023erp}, this study focuses on the variations in EEG and EM signals across different target categories to assess the feasibility of distinguishing between signals evoked by different target classes.

\subsubsection{EEG Signals Analysis}
    Considering the temporal dynamics and spatial topological characteristics of raw EEG signals, we apply the non-parametric permutation cluster repeated measures ANOVA test \cite{gramfort2014mne} to compare data from all subjects in Task A. The results reveal four significant spatiotemporal clusters ($p<0.05$) between EEG signals across target-1 and target-2 classes (see Fig. \ref{EEG difference}). Each subfigure in Fig. \ref{EEG difference} represents a spatiotemporal cluster, where each cluster contains EEG sampling points that exhibit significant differences across two classes and are adjacent in both temporal and spatial (channel) dimensions. 

    Figure \ref{EEG difference} shows significant differences in EEG signals between the two target classes across multiple time intervals and brain regions. In the 24 to 280 ms interval, significant differences are observed across the entire brain region (see Fig. \ref{EEG difference}(a)). Figure \ref{EEG difference}(b) highlights differences in the occipital and parietal lobes over the span of 276 to 400 ms, associated with the N200 component. In the 428 to 612 ms interval where the P300 component appears, there are significant differences in the occipital and parietal lobes (see \ref{EEG difference}(c)). Lastly, Fig \ref{EEG difference}(d) demonstrates significant differences in the occipital and parietal lobes during the 636 to 928 ms interval. These results indicate that the EEG signal evoked by the targets from the two classes has significant spatial and temporal differences. Spatially, these differences are concentrated in the parietal and occipital lobes. Temporally, the differences manifest during both early and late phases of stimulus-evoked ERPs, as well as the period where N200 and P300 components appear.

\subsubsection{EM Signals Analysis}
    Since the components of EM signals exhibit minimal spatial correlation compared with EEG signals, we analyze the pupil area (see Fig \ref{EM difference}(a)), horizontal position (see Fig \ref{EM difference}(b)), and vertical position (see Fig \ref{EM difference}(c)) of EM signals separately. Due to the high temporal synchronization of binocular movements, we draw the grand-averaged EM responses of the left eye from all subjects in Task A.    
    
    \begin{figure*}[ht]
        \centering
        \subfigure[Pupil area]{
			\includegraphics[width=0.25\linewidth]{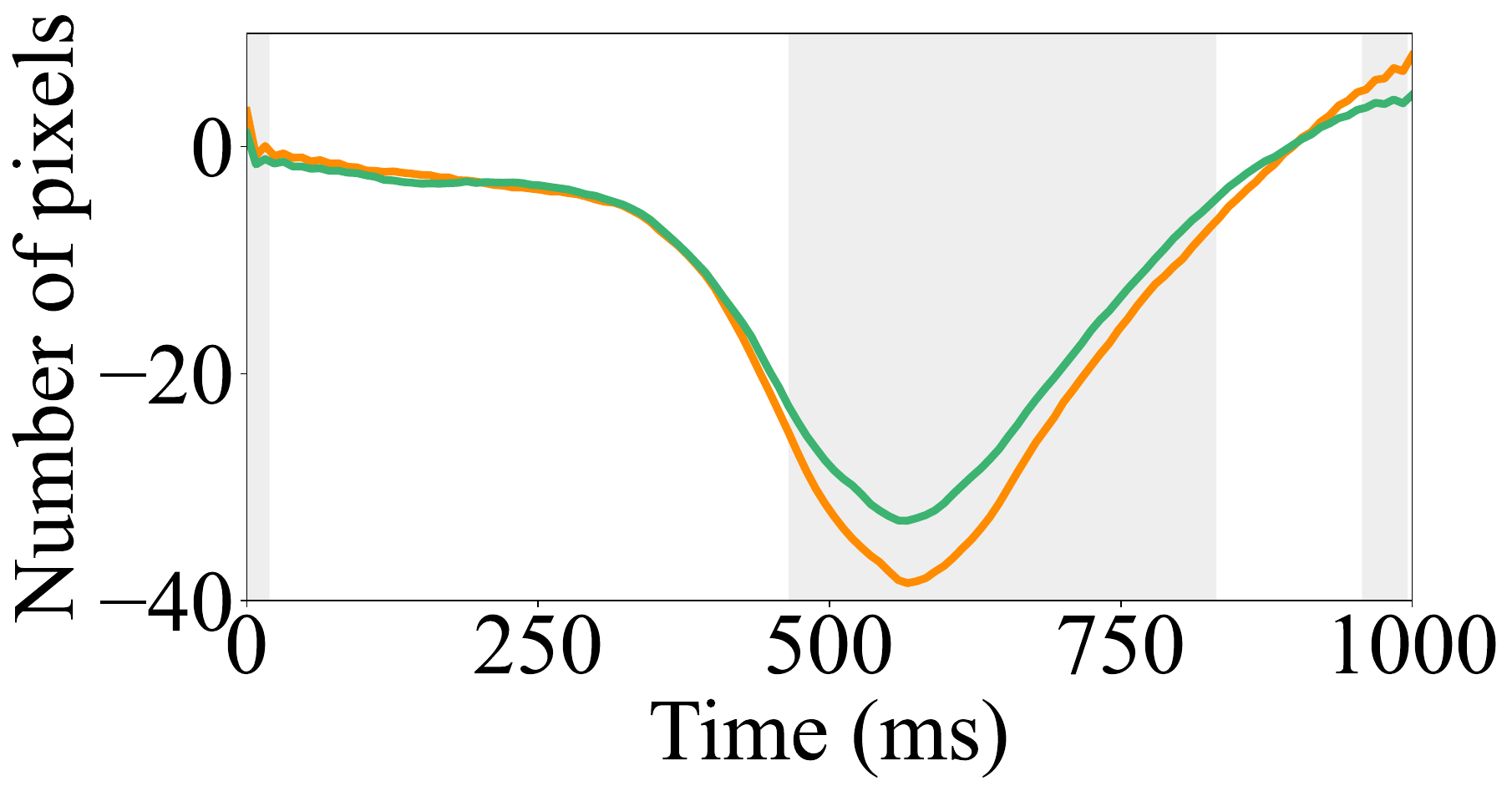}
		}\hspace{0.15cm}
		\subfigure[Horizontal position]{
			\includegraphics[width=0.25\linewidth]{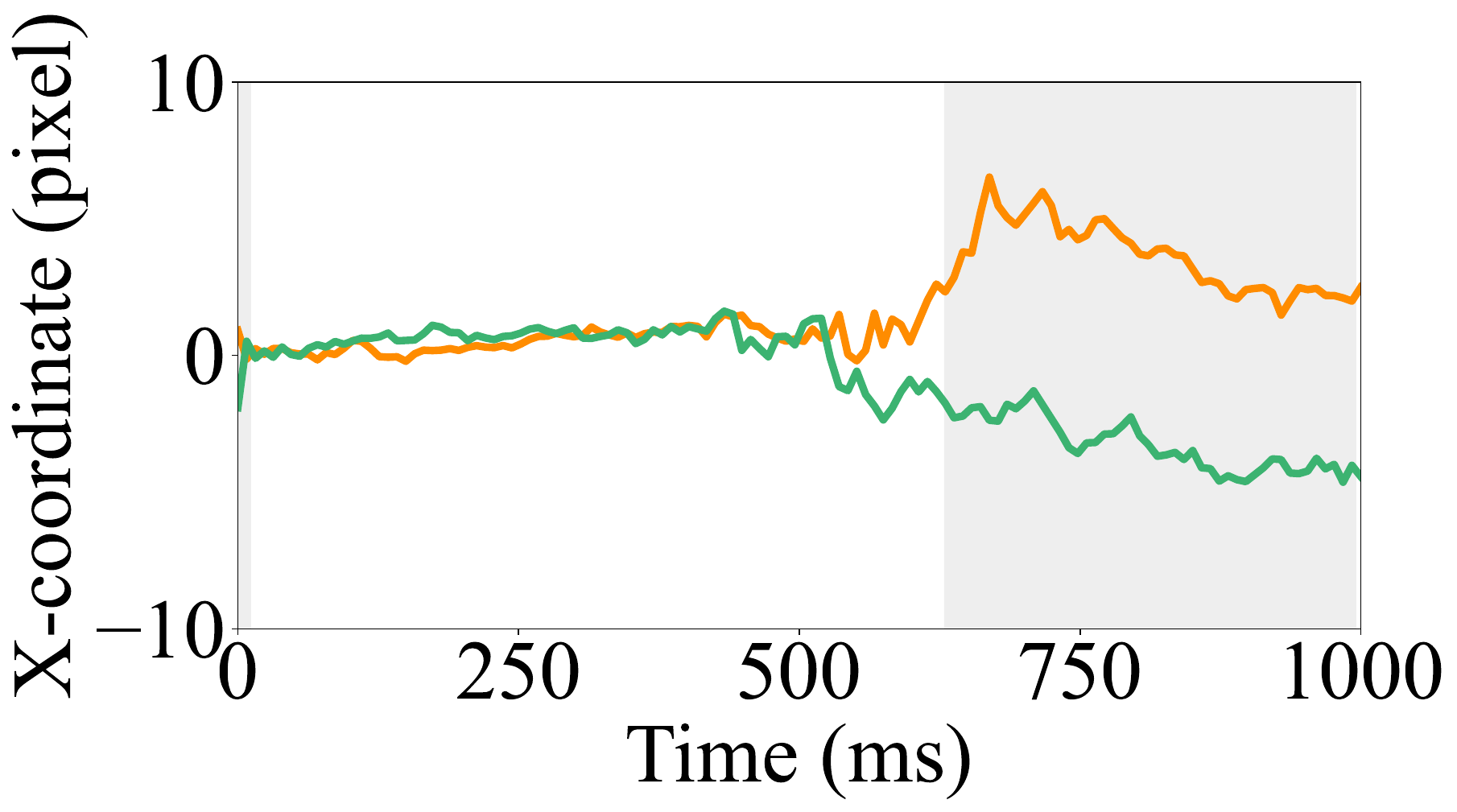}
		}\hspace{0.15cm}
        \subfigure[Vertical position]{
			\includegraphics[width=0.25\linewidth]{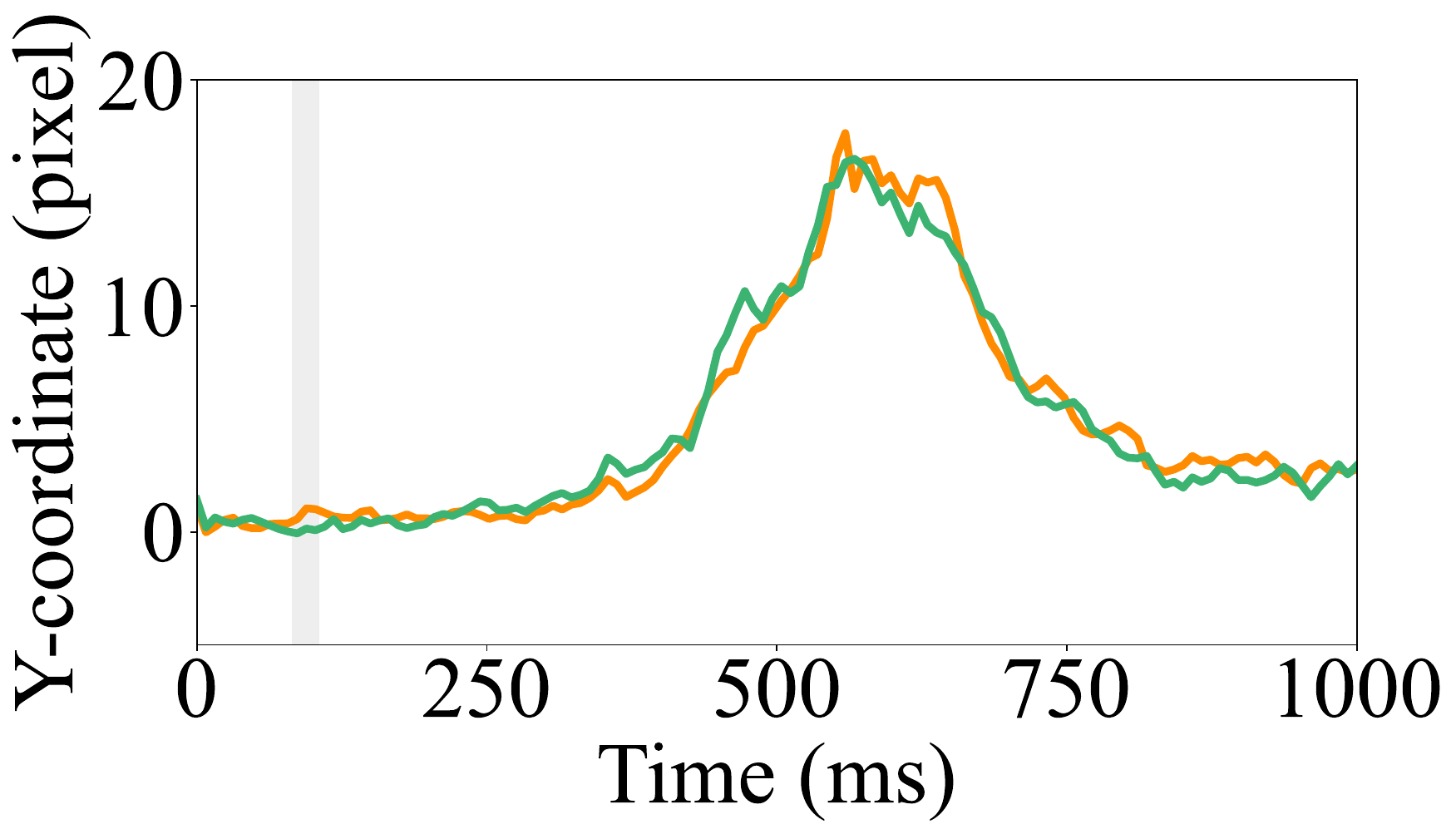}
		}\hspace{0.2cm}
        \subfigure{
			\includegraphics[width=0.12\linewidth]{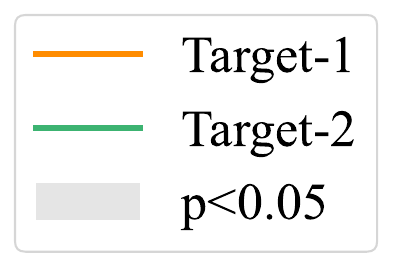}
		}
		\caption{The one-way ANOVA analysis of grand-averaged target-1 and target-2 EM responses from all subjects in Task A. Subfigures (a)-(c) represent the evoked EM modality responses including pupil area and the horizontal and vertical gaze position from the left eye. The Y-axis in (a) represents the area of the pupil measured by the number of pixels on the screen, while the Y-axis in (b) and (c) denote the position of the gaze point with respect to the baseline in pixels. The light orange shaded regions indicate significant differences ($p<0.05$) among the two target classes at that moment.}
        \label{EM difference}
    \end{figure*}

    One-way ANOVA is employed to evaluate temporal differences in EM components between the two target classes. The analysis shows significant temporal differences in pupil area and horizontal position between the two target classes ($p<0.05$), while vertical position exhibits no notable variations. The pupil area displays significant differences primarily during 468-820 ms post-stimulus onset ($p<0.05$), and horizontal position differences occur during 632-992 ms ($p<0.05$). These results demonstrate significant differences in pupil area and horizontal position of EM signals evoked by different classes of targets, indicating the feasibility of introducing EM modality to enhance multi-class target RSVP decoding.

\subsection{Saliency Map Analysis}
    We employ the saliency map method \cite{simonyan2013deep} to analyze the relative importance of spatiotemporal components in EEG and EM signals within MTREE-Net. Saliency maps are generated by computing gradients of output with respect to input features, highlighting signal regions that significantly influence model predictions. In this experiment, the saliency maps of the EEG and EM input pairs are calculated for samples of two target classes during testing. Subsequently, the saliency maps of all subjects are averaged. For EEG modality, We visualize saliency maps in both spatial (channel) and temporal dimensions. The channel and temporal importance scores are obtained by accumulating saliency map values along channel and time dimensions, respectively (see Fig. \ref{saliency map}(a)). For the EM modality, we generate saliency maps for left eye movements across both component and temporal dimensions. Component and temporal importance scores are computed by accumulating saliency map values along component-wise and temporal axes (see Fig. \ref{saliency map}(b)).

    \begin{figure}[!htbp]
        \centering
        \subfigure[EEG modality]{
			\includegraphics[width=0.98\linewidth]{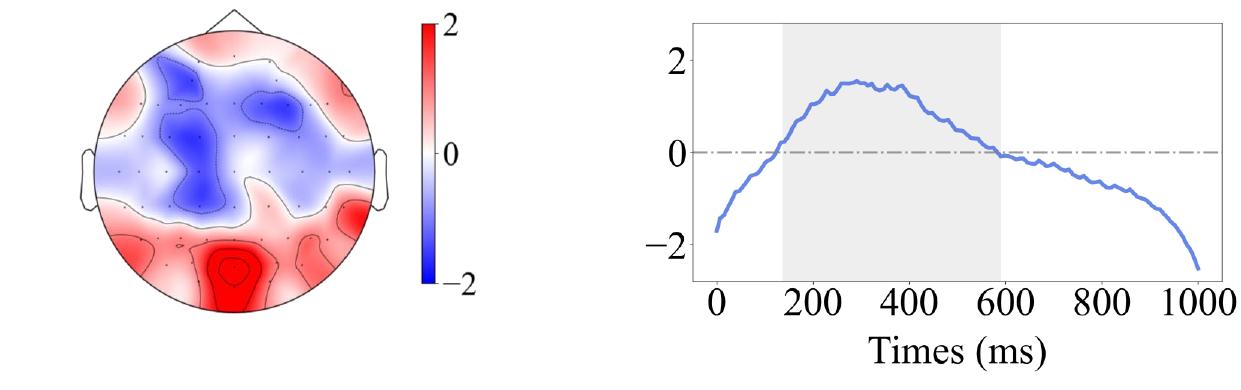}
		}
		\subfigure[EM modality]{
			\includegraphics[width=0.98\linewidth]{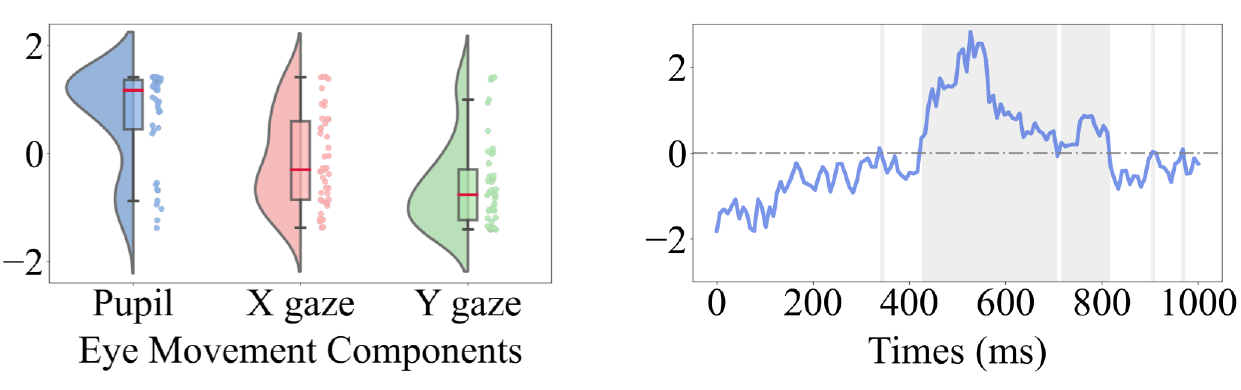}
		}
		\caption{The averaged saliency maps of (a) EEG modality and (b) EM modality across all subjects in Task A. For EEG modality, the left displays the saliency maps accumulated in the channel dimension, while the right shows the saliency maps accumulated in the time dimension. For EM modality, the left represents component-wise accumulated saliency maps through violin plots and box plots with subject-specific points, and the right shows temporal aggregation of saliency maps. Light orange regions in temporal plots indicate time points with above-average values.}
        \label{saliency map}
    \end{figure}

    Fig. \ref{saliency map}(a) displays the normalized importance scores for EEG channels and time points. The saliency analysis reveals that target classification is primarily influenced by brain activities in parietal and occipital regions, as well as within the time range of 148-585 ms. These findings are consistent with our previous analysis in Fig. \ref{EEG difference}, where the differences between EEG signals evoked by different targets are mainly concentrated in the parietal and occipital lobes spatially, and are partially reflected during the time periods of N200 and P300 components.  Fig. \ref{saliency map}(b) displays the normalized importance scores for EM components and time points. The model's classification is most influenced by pupil area, followed by the horizontal position, and then the vertical position. This aligns with the findings in Fig. \ref{EM difference} where the vertical position of EM signals shows negligible significant differences between the two target classes. Temporally, the higher classification influence of EM signals occurs during 437-812 ms, corresponding to the period where Fig. \ref{EM difference} demonstrates significant differences in pupil area and horizontal position of EM signals evoked by two target classes. These results demonstrate that our model effectively captures discriminative spatiotemporal patterns across both modalities. In EEG modality, the model focuses on the parietal and occipital regions as well as the N200 and P300 components. In EM modality, it utilizes variations in pupil area and horizontal position. The combined features from both modalities enable efficient multi-class target RSVP decoding.  

\section{Discussion}

\subsection{Performance of Each Component in The EM Modality}
    The EM modality used in our method comprises three components: horizontal position (X), vertical position (Y), and pupil area. The saliency map analyses indicate that the model's classification is most influenced by pupil area, followed by horizontal position and vertical position. To further analyze the contribution of each EM component to MTREE-Net's performance, we conduct experiments by combining each EM component with EEG data for multi-class RSVP decoding using MTREE-Net.

    \setlength{\tabcolsep}{2.1mm}{
    \begin{table}[htbp]
        \begin{threeparttable}
            \small
            \renewcommand\arraystretch{1.3}
            \centering
            \caption{Balanced-accuracy ($\%$) of MTREE-Net with different EM components (mean).}
            \label{EM parts}
            \begin{tabular}{llll}
            \toprule[1.2pt]
                \multirow{2}{*}{\textbf{ EM components}}&\multicolumn{3}{c}{\textbf{ Task}}\\
                \cmidrule(lr){2-4}
                & \textbf{ A} & \textbf{ B} & \textbf{ C}\\
              \hline
               Only horizontal (X) position & $67.92^{\star\star\star} $ & $73.31^{\star\star\star}  $&  $72.10^{\star\star\star} $  \\
                Only vertical (Y) position & $67.60^{\star\star\star} $&  $73.45^{\star\star\star}$ & $71.86^{\star\star\star} $  \\
                Only pupil area & $ \underline{68.24}^{\star\star}$& $\underline{74.00}^{\star}  $&$\underline{72.44}^{\star} $ \\
                All components (ours) &  $\mathbf{68.96} $ &   $\mathbf{74.42} $ &  $\mathbf{73.05} $ \\
              \bottomrule[1.2pt]
            \end{tabular}
            \begin{tablenotes}
                \item \footnotesize The asterisks in the table indicate a significant difference between the performance of MTREE-Net using different EM components by paired t-tests ($\star p<0.05,\star\star p<0.01,\star\star\star p<0.001$).
            \end{tablenotes}
        \end{threeparttable}
    \end{table}} 

    The results are presented in Table \ref{EM parts}. A two-way repeated measures ANOVA indicates significant main effects for both the EM components used in MTREE-Net ($p<0.001$) and the tasks ($p<0.001$), with no significant interaction effects between these factors ($p=0.25$). Post-hoc tests reveal that, in each task, MTREE-Net using all EM components significantly outperforms the models using any single EM component (all: $p<0.05$). Furthermore, the post-hoc analysis shows that the BA of MTREE-Net using only pupil area data is significantly higher than that of the model using the X position across all three tasks (all: $p<0.01$) and higher than the model using the Y position. In contrast, MTREE-Net using only the X position achieves the lowest BA among the models using a single EM component.

    These results demonstrate that incorporating all EM components into the proposed method significantly enhances multi-class RSVP decoding performance. Among models using a single EM component, the model relying on pupil area achieves the highest performance, supporting the observation in Fig. \ref{saliency map} that pupil area has the most substantial impact on classification predictions. Conversely, the model using only the Y position exhibits the lowest performance, further validating the finding in Fig. \ref{EM difference} that vertical position signals show minimal significant differences between the two target classes.

\subsection{Comparison of Different Complementary Direction}
    The DCM effectively improves feature differentiation across categories for each modality and enhances classification performance by employing bidirectional feature complementarity between EEG and EM modalities. To further investigate the impact of complementary directions in the dual-complementary module (DCM), we conduct experiments with three configurations in MTREE-Net: EEG-to-EM, EM-to-EEG, and dual complementarity. The EEG-to-EM configuration enhances EM representations using EEG features, as formulated in Equ. (\ref{EEG to EM}). Conversely, the EM-to-EEG configuration improves EEG representations using features extracted from EM signals, as outlined in Equ. (\ref{EM to EEG}). The dual complementarity configuration combines both approaches, representing the full implementation of the DCM.

    \setlength{\tabcolsep}{2.1mm}{
    \begin{table}[htbp]
        \begin{threeparttable}
            \small
            \renewcommand\arraystretch{1.3}
            \centering
            \caption{Balanced-accuracy ($\%$) of MTREE-Net with different complementary directions (mean).}
            \label{complementary direction}
            \begin{tabular}{llll}
            \toprule[1.2pt]
                \multirow{2}{*}{\textbf{ Complementary Direction}}&\multicolumn{3}{c}{\textbf{ Task}}\\
                \cmidrule(lr){2-4}
                & \textbf{ A} & \textbf{ B} & \textbf{ C}\\
              \hline
               EEG to EM & $67.93^{\star\star\star} $ & $73.57^{\star\star\star}  $&  $72.09^{\star\star\star} $  \\
                EM to EEG & $\underline{68.19}^{\star\star\star} $&  $\underline{73.87}^{\star\star\star}$ & $\underline{72.43}^{\star\star\star} $  \\
                Dual complementarity (ours) &  $\mathbf{68.96} $ &   $\mathbf{74.42} $ &  $\mathbf{73.05} $ \\
              \bottomrule[1.2pt]
            \end{tabular}
            \begin{tablenotes}
                \item \footnotesize The asterisks in the table indicate a significant difference between the performance of MTREE-Net and the ablation methods using different complementary directions by paired t-tests ($\star p<0.05,\star\star p<0.01,\star\star\star p<0.001$).
            \end{tablenotes}
        \end{threeparttable}
    \end{table}}

    The results are presented in Table \ref{complementary direction}. A two-way repeated measures ANOVA reveals significant main effects for both complementary direction ($p<0.001$) and task ($p<0.001$), with no significant interaction effects between these factors ($p=0.59$). Post-hoc analysis indicates that the dual complementarity configuration achieves the highest BA in all three tasks compared to the EEG-to-EM and EM-to-EEG configurations (all: $p<0.001$). This improvement is attributed to the complementarity between the two modalities, which simultaneously enhances the distinction between two uni-modal features across different categories. Moreover, the EM-to-EEG configuration slightly outperforms the EEG-to-EM configuration, where the EM is the weaker modality and EEG is the stronger modality in terms of classification performance. This result indicates that using features from the weaker modality (EM) to enhance the stronger modality (EEG) is more effective in improving the performance of EEG and EM feature fusion, compared to using features from the stronger modality to supplement the weaker modality.

\subsection{Limitations and Future Work}
     In our three multi-class target RSVP experiments, we utilize the typical RSVP presentation speed of 10 Hz without considering other presentation rates. Future work can explore the effects of different RSVP presentation rates on the induced EEG and eye movement signals, and the decoding performance of the model. Additionally, our proposed approach requires complete multi-modal samples, which may be challenging to guarantee in real-world applications. Thus, future work can focus on developing a model that can perform effectively on both complete multi-modal data and incomplete data with partially missing modalities.

\section{Conclusion}
    In this study, we explore the fusion of EEG and EM signals to enhance multi-class target RSVP decoding performance. We design three independent multi-class target RSVP tasks and build an open-source dataset that includes both EEG and EM signals. We then propose the Multi-class Target RSVP EEG and EM fusion Network (MTREE-Net) to enhance multi-class RSVP decoding. MTREE-Net employs a dual-stream feature extractor to extract multi-scale EEG features and compress the features of EM components. A dual-complementary module is introduced to simultaneously improve the differentiation of uni-modal features across classes. We also analyze the contribution of each modality to the classification results in theory and introduce a modal contribution ratio to guide the optimization of modal reweighting fusion. Furthermore, a hierarchical self-distillation module is designed to reduce the misclassification of non-target samples by transferring knowledge from the binary classifier to guide the triplet classifier. Extensive experiments on our open-source dataset demonstrate that MTREE-Net significantly outperforms existing EEG decoding and EEG-EM fusion methods. The experiments also prove the feasibility of introducing EM signals to enhance multi-class RSVP decoding and the effectiveness of each proposed module. Our collected open-source dataset and proposed MTREE-Net provide a promising framework for developing multi-class RSVP-BCI systems in practical applications.
    
\bibliographystyle{elsarticle-template-num} 
\bibliography{elsarticle-template-num}

\end{document}